\documentclass[letterpaper]{article}
\usepackage{graphicx}
\usepackage{amssymb}

\DeclareMathAlphabet{\bm}{OML}{cmm}{b}{it}
\newcommand{\ket}[1]{| #1 \rangle} 
\newcommand{\bra}[1]{\langle #1 |} 

\newcommand{\rom}[1]{\mathrm{#1}}
\newtheorem{thm}{Theorem}
\newtheorem{lem}[thm]{Lemma}
\newtheorem{definition}[thm]{Definition}

\title{Noise Tolerance of the BB84 Protocol with Random Privacy
Amplification\thanks{Part of this paper 
will be presented
in the 2005 International Symposium on Information Theory,
Adelaide Convention Centre, Adelaide, Australia,
4--9 September, 2005.}}

\author{
Shun Watanabe 
\and
Ryutaroh Matsumoto \and
Tomohiko Uyematsu \and \\
Department of Communications and Integrated Systems, \\
 Tokyo
Institute of Technology, Tokyo 152-8552, Japan
}
\date{June 24, 2005}
\begin{document}
\maketitle  


\begin{abstract}
This paper shows that the BB84 protocol with random privacy amplification is
secure with a higher key rate than Mayers' estimate with
the same error rate.
Consequently, the tolerable error rate of this
protocol is increased from 7.5 \% to 11 \%.
We also extend this method to the case of estimating
error rates separately in each basis, which enables us to 
securely share a longer key. 

\end{abstract}


\section{Introduction}

The BB84 protocol is the first quantum key distribution (QKD) protocol,
which was proposed by Bennett and Brassard in 1984 \cite{bennett84}.
Unlike conventional cryptographies that rely on the 
conjectured difficulty of computing certain functions,
the security of QKD is guaranteed by the postulate of quantum mechanics.
In the BB84 protocol, the participants (Alice and Bob) agree
on a secret key about which any eavesdropper (Eve) can
obtain little information.
The security proof of this protocol against
arbitrary eavesdropping strategies was first proved by Mayers \cite{mayers01},
and a simple proof was later shown by Shor and Preskill \cite{shor00}.
Later, many security analyses are 
studied \cite{hamada04,gottesman03,wang04a,lo00,gottesman00}.

In the BB84 protocol, two linear codes $C_1$ and $C_2$ 
are employed to
share a secret key.
$C_1$ is used for error correction, and $C_2$ is used for
privacy amplification.
Error correction is performed to share the same key, which is not
necessarily secret.
Privacy amplification is performed to extract a shorter secret key.
To share the same secret key,
$C_2$ must be a subcode of $C_1$, and the decoding error
probability of $C_1$ and $C_2^\bot$ 
as a CSS code must be small.
For the key distribution protocol to be practical, we require the linear
code $C_1$ to be efficiently decodeable.
However, it is difficult to find a pair of linear codes $C_1$ and $C_2$ that
satisfy these conditions.
Because we do not have to decode $C_2^\bot$, it is sufficient
that $C_2^\bot$ is a randomly chosen code whose decoding error probability
is small with the maximum likelihood decoding.
Mayers showed that if one determines $C_1$ and
chooses $C_2$ with rate $\rom{H}(2p)$
at random from subcodes of $C_1$,
the minimum Hamming weight of
$C_2^\bot \backslash C_1^\bot$ is greater than $p n$
with high probability,
where $p$ is an estimated error rate and 
$\rom{H}(\cdot)$ is the binary entropy function
\cite[Lemma 4]{mayers01}.
Consequently, the decoding error probability of $C_2^\bot$ 
as a part of a CSS code is small.
With this method, 
we can share a key with the key rate  $1-\rom{H}(p)-\rom{H}(2p)$, and
the protocol can tolerate 
error rates up to  7.5\%.
In this paper, we call the random privacy amplification
as the method such that one chooses $C_2$ at random from  
subcodes of a fixed code $C_1$ and
performs the privacy amplification by $C_2$.

However, by evaluating directly the decoding error probability of $C_2^\bot$ instead of
the minimum Hamming weight, we can decrease the rate of $C_2$ while
maintaining the security of the protocol.
This paper shows that when one
chooses $C_2$ with rate $\rom{H}(p)$ at random from
subcodes of $C_1$, the decoding
error probability of $C_2^\bot$ 
as a part of a CSS code is exponentially small with high probability.
Consequently,
when we choose a code $C_2$ at random in the BB84 protocol, 
according to our evaluation of decoding error probability, we can share
a key with
the key rate $1-2\rom{H}(p)$ and the protocol can tolerate  
error rates up to 11\%. 

To share a key more efficiently,
it is known that we should estimate error rates, $p_0$ and $p_1$,
separately in two basis 
$\{\ket{0},\ket{1}\}$ and
$\{\frac{\ket{0}+\ket{1}}{\sqrt{2}},\frac{\ket{0}-\ket{1}}{\sqrt{2}}\}$
\cite{hamada04,gottesman03,wang04a,lo00}.
This paper also shows that
if one chooses $C_2$ of a rate 
$\frac{\rom{H}(p_0)+\rom{H}(p_1)}{2}$
at random from subcodes of $C_1$,
the decoding error probability of
$C_2^\bot$ 
as a part of a CSS code is exponentially small with high probability,
which is proved by an analogue method in \cite[Appendix B]{hamada04}.

It is also known that QKD protocols with two-way classical communications
can tolerate higher error rate than QKD protocols with one-way classical
communications.
The tolerable error rates are
18.9\% in \cite{gottesman03}, 20\% in \cite{chau02},
and 26\% in \cite{wang04b}.
Our result on random privacy amplification is also
applicable for those protocols,
because they perform error correction and
privacy amplification after reducing the error rate
with two-way classical communications.

We stress that our result is different from 
previously known results.
In \cite[Lemma 4]{mayers01},
it is proved that if we fix $C_1$ of rate $1-\rom{H}(p)$ and
choose its subcode $C_2$ of rate $\rom{H}(2p)$ at random,
the BB84 protocol is secure, which means that
the BB84 protocol with random privacy amplification
can tolerate the error rate of 7.5\%.
In \cite{shor00}, the authors state that there exists a pair of
$C_1$ and $C_2$ by which
the BB84 protocol can tolerate the error rate of 11\%.
However, one cannot guarantee that $C_1$ is efficiently decodeable. 
They also cite \cite[Lemma 4]{mayers01} in order
to show that we can securely choose a random subcode
$C_2$ of an efficiently decodeable code
$C_1$. However, it is not clarified in \cite{shor00} 
whether or not  the BB84 protocol with random privacy amplification
can tolerate the 11\% error rate.
Other previous papers \cite{hamada04,gottesman03,wang04a,lo00,gottesman00}
are based on the result in \cite{shor00}.
Thus nobody has proved that the BB84 protocol with
random privacy amplification can tolerate the 11\% error rate.
We also stress that the random hashing method cannot be directly
applied to the security proof of
the BB84 protocol with random privacy amplification as used in \cite{lo01},
because a fixed $C_1$ and the condition $C_2 \subset C_1$
decrease the randomness of hashing.
Application of the random hashing to a security proof of the
random privacy amplification requires a careful argument similar
to Section \ref{random-privacy} of this paper.

This paper is organized as follows.
In Section \ref{bb84-protocol},
we introduce the BB84 protocol, and 
present the required conditions on $C_1$ and $C_2$.
We also relate those conditions to the security
of the BB84 protocol quantitatively.
In Section \ref{random-privacy},
the main theorem is proved.
Concluding remarks are given in Section \ref{conclusion}.

\section{The BB84 protocol}
\label{bb84-protocol}

We consider the following BB84 protocol
modified from \cite{shor00}.
As shown in \cite{hamada04,gottesman03,wang04a,lo00}, we estimate error rates
separately in two basis
$\{\ket{0},\ket{1}\}$ and
$\{\frac{\ket{0}+\ket{1}}{\sqrt{2}}, \frac{\ket{0}-\ket{1}}{\sqrt{2}}\}$.
As is also mentioned in \cite{shor00,hamada04,gottesman03},
Alice and Bob agree on a random permutation $\pi$ after
transmission of the qubits and use the linear codes 
scrambled by $\pi$,
where $\pi$ scrambles the $n$-bit vector within first $\frac{n}{2}$ bits
and latter $\frac{n}{2}$ bits respectively, i.e.,
$\pi:(x_1,\cdots,x_{\frac{n}{2}},y_1,\cdots,y_{\frac{n}{2}}) \mapsto
(x_{\pi_1(1)},\cdots,x_{\pi_1(\frac{n}{2})},y_{\pi_2(1)},$
$\cdots,y_{\pi_2(\frac{n}{2})})$,
and $\pi_1,\pi_2 \in S_{\frac{n}{2}}$ are permutations on $\{1,\cdots,\frac{n}{2}\}$.
By this procedure,
we can securely share a key against general eavesdropping attacks 
with a linear code whose decoding error probability
as a part of a CSS code
is small over a BSC (Binary Symmetric Channel) 
\cite[Lemmas 2, 3]{gottesman03}.

\subsection{The BB84 protocol}

\renewcommand{\labelenumi}{(\theenumi)}
\begin{enumerate}
\item Alice randomly select $(4+\theta)n$-bit strings $\bm{k}$ and $\bm{a}$, and
chooses a random permutation $\pi$.
\item Alice repeats the following procedures for $1 \le i \le (4+\theta)n$.
If $a_i=0$, she creates either state $\ket{0}$ for $k_i=0$ or
$\ket{1}$ for $k_i=1$.
If $a_i=1$, she creates either state $\ket{+}$ for $k_i=0$ or
$\ket{-}$ for $k_i=1$. 
We represent prepared states as $\ket{\varphi_i}$, where
$\varphi_i \in \{0,1,+,-\}$, $\ket{+}=\frac{\ket{0}+\ket{1}}{2}$,
$\ket{-}=\frac{\ket{0}-\ket{1}}{2}$.
\item Alice sends the resulting $(4+\theta)n$ qubits
$\ket{\varphi_1} \otimes \cdots \otimes \ket{\varphi_{(4+\theta)n}}$
to Bob.
\item Bob receives the $(4+\theta)n$ qubits
$\ket{\tilde{\varphi}_1} \otimes \cdots \otimes \ket{\tilde{\varphi}_{(4+\theta)n}}$.
\item Bob randomly select $(4+\theta)n$-bit string $\bm{b}$.
\item Bob repeats the following procedures
for $1 \le i \le (4+\theta)n$.
If $b_i=0$, he measures $\ket{\tilde{\varphi}_i}$ with $\sigma_z$.
If $b_i=1$, he measures $\ket{\tilde{\varphi}_i}$ with $\sigma_x$.
Then, measurement result, $+1$ and $-1$, corresponds to $\tilde{k}_i=0$
and $\tilde{k}_i=1$, respectively.
After these procedures Bob will obtain
$\tilde{\bm{k}}=(\tilde{k}_1,\cdots,\tilde{k}_{(4+\theta)n})$.
\item Alice announces $\bm{a}$ and $\pi$.
\item If $a_i \neq b_i$, Alice and Bob discard $i$-th bit of
$\bm{k}$ and $\tilde{\bm{k}}$.
With high probability, at least $2n$ bits remain, and
there are at least $n$ bits where $a_i=b_i=0$, and there are at least
$n$ bits where $a_i=b_i=1$ (if not, abort the protocol).
\item Alice chooses $n$ bits where $a_i=b_i=0$, and
divides them into two $\frac{n}{2}$-bit strings, $\bm{c}_0$ and $\bm{d}_0$.
She chooses $n$ bits where $a_i=b_i=1$, and divides them into
two $\frac{n}{2}$-bit strings, $\bm{c}_1$ and $\bm{d}_1$.
Alice announces which bits are $\bm{c}_0$, $\bm{d}_0$, $\bm{c}_1$, $\bm{d}_1$.
Then, Bob will obtain   
$\tilde{\bm{c}}_0=\bm{c}_0+\bm{e}_0$, $\tilde{\bm{c}}_1=\bm{c}_1+\bm{e}_1$
$\tilde{\bm{d}}_0=\bm{d}_0+\bm{f}_0$, $\tilde{\bm{d}}_1=\bm{d}_1+\bm{f}_1$, 
where $\bm{e}_0$, $\bm{e}_1$, $\bm{f}_0$, $\bm{f}_1$ are errors 
caused by eavesdropping and channel noise.
\label{divide}
\item Alice and Bob compare $\bm{d}_0$ with $\tilde{\bm{d}}_0$
and $\bm{d}_1$ with $\tilde{\bm{d}}_1$, then they obtain
$\bm{f}_0$ and $\bm{f}_1$.
From $\bm{f}_0,\bm{f}_1$,
Alice and Bob choose a pair of linear codes $C_1$ and $C_2$
that satisfy the conditions (\ref{con-a})--(\ref{con-c})
in Section \ref{security-condition}.
If there exists no such a pair of linear codes, then
they abort the protocol.
\label{step10}
\item Alice chooses a random codeword $\bm{v}$ from $\pi(C_1)$ whose
length is $n$,
where $\pi(C_1)$ is a code that all codewords in $C_1$ are
permuted by $\pi$.
She sends $\bm{x}=\bm{v}+\bm{c}$ with a public classical channel,
where $\bm{c}$ is a concatenation of $\bm{c}_0$ and $\bm{c}_1$.
\item Bob receives $\bm{x}=\bm{v}+\bm{c}$ and subtracts $\tilde{\bm{c}}$
from it.
Then, he corrects $\bm{v}+\bm{e}$ to a codeword $\hat{\bm{v}}$ in
$\pi(C_1)$, where $\bm{e}$ is a concatenation of $\bm{e}_0$ and $\bm{e}_1$,
and $\tilde{\bm{c}}$ is a concatenation of $\tilde{\bm{c}}_0$ and $\tilde{\bm{c}}_1$.
\label{error-correction}
\item Alice uses the coset of $\bm{v}+\pi(C_2)$ as a key, and 
Bob uses the coset of $\hat{\bm{v}}+\pi(C_2)$ as a key.
\end{enumerate}

\subsection{Security of the protocol}
\label{security-condition}

The security of the BB84 protocol can be proved by
showing the security of the CSS code protocol 
(QKD using a CSS code) \cite{hamada04}.
Maintaining the security, 
the BB84 protocol is related with the CSS code protocol.
This kind of technique was first used in \cite{shor00},
in which the BB84 protocol is related to the EPP
(Entanglement Purification Protocol) protocol. 
If $C_1$ and $C_2$ satisfy the following three conditions,
then a shared key is secure against general eavesdropping attacks.

\renewcommand{\theenumi}{\alph{enumi}}
\begin{enumerate}
\item $C_2 \subset C_1$
\label{con-a}
\item 
\label{con-b}
If the crossover probability of first $\frac{n}{2}$ bits of the BSC are
smaller than or equal to $p_0$ and the crossover probability of latter 
$\frac{n}{2}$ bits are smaller than or equal to $p_1$,
then the decoding error probability of $C_1$ 
as a part of a CSS code over the BSC, whose formal definition is
given in Definition \ref{def1}, is
smaller than or equal to $\epsilon$. 
\item 
\label{con-c}
If the crossover probability of first $\frac{n}{2}$ bits of the BSC are
smaller than or equal to $p_1$ and the crossover probability of latter 
$\frac{n}{2}$ bits are smaller than or equal to $p_0$,
then the decoding error probability of $C_2^\bot$ 
as a part of a CSS code over the BSC is
smaller than or equal to $\epsilon$.
\end{enumerate}

We set $p_0$ and $p_1$ to
$p_0=Q_{\bm{f}_0}(1)+\delta$ and $p_1=Q_{\bm{f}_1}(1)+\delta$
in step (\ref{step10}), where
$Q_{\bm{f}_0}$, $Q_{\bm{f}_1}$ are the types
of $\bm{f}_0$,$\bm{f}_1$
(refer to \cite{csiszar-korner81} for the definition of the type), and
$\delta$ and $\epsilon$ are sufficiently small positive numbers.
Throughout this paper, we assume $p_0 < \frac{1}{2}$ 
and $p_1 < \frac{1}{2}$.

We stress that
the decoding error probability of $C_1$ and $C_2^\bot$ have
to be small over any BSC with
crossover probability below $p_0$ and 
crossover probability below $p_1$, 
instead of a single BSC with crossover probabilities $p_0$ and $p_1$.
The necessity of such a requirement on decoding error probability
is already observed in \cite{hamada04}, \cite[Proof of Lemma 3]{gottesman03}.
 
The security of the BB84 protocol
is usually evaluated by the mutual information
between a shared key and Eve's accessible information.
In order to implement the BB84 protocol,
the designer of the system has to find 
a pair of linear codes by which 
the mutual information between a shared key
and Eve's accessible information
is smaller than an acceptable level.
To find such a pair of linear codes,
we need a criterion according to which
we can distinguish whether a particular pair of linear
codes makes the mutual information smaller than 
an acceptable level.

In the security proof of \cite{gottesman03}, it is  
proved that the security of the BB84 protocol against general 
eavesdropping attack is reduced to the security against
uncorrelated Pauli attacks (Eve applies a random Pauli operator
independently on each qubit sent through the channel).
However, only a asymptotic upper bound on the
mutual information is proved,
and the authors do not present
a sufficient condition for low mutual information
on a pair of linear codes 
of a finite code length.

In the security proof of \cite{hamada04},
Hamada presents a condition on a pair of linear codes \cite[Corollary
2]{hamada04},
and proves that the mutual information is upper bounded quantitatively
by a pair of linear codes satisfying that condition.
However, that condition does not aid choosing a
suitable linear code, because we cannot easily decide
whether a particular code satisfies it.

By upper bounding the mutual information by a function
of the decoding error probability $\epsilon$,
we can find a pair of linear codes that
makes the mutual information smaller than an acceptable
level according to the conditions (\ref{con-a})--(\ref{con-c}).
Because evaluating an upper bound on the
 decoding error probability of a code
is not difficult,
the conditions (\ref{con-a})--(\ref{con-c})
on a pair of linear codes are practically useful.
The following theorem gives an upper bound on the 
mutual information as a function of the decoding 
error probability $\epsilon$.

\begin{thm}
\label{theorem-security}
If we use linear codes $C_1$ and $C_2$ that satisfy 
the conditions (\ref{con-a})--(\ref{con-c}) in the 
BB84 protocol,
then the mutual information between a shared key and 
Eve's accessible information
(including messages exchanged over the classical channel) is upper bounded by
\begin{eqnarray*}
&& \rom{I}(\bm{U};\bm{E},\bm{S}) \\
&\le& \rom{H}\left(
2(\frac{n}{2}+1)^2 \epsilon + 2\exp \left\{-\Theta(\delta^2 n)\right\}
\right)
+ 4n(\frac{n}{2}+1)^2 \epsilon + 4n \exp\left\{-\Theta(\delta^2 n)\right\},
\end{eqnarray*}
where the base of $\exp(\cdot)$ is $2$, $\Theta(\delta^2 n)$ is given by
\begin{eqnarray*}
\Theta(\delta^2 n) = \frac{\delta^2}{4 \ln 2}n - 2 \log (n+1) -2,
\end{eqnarray*}
$\bm{S}$ denotes the random variable
of the information transmitted through the public classical channel,
$\bm{U}$ denotes the random variable of a shared key, i.e.,
the coset of $\bm{v}+\pi(C_2)$,
and $\bm{E}$ denotes the random variable of Eve's 
eavesdropping result from transmitted qubits.
\end{thm}
This theorem is proved in Appendix \ref{appendix-security}.
Note that the upper bound of the mutual 
information is valid for finite $n$.


\section{Random privacy amplification}
\label{random-privacy}

To implement the BB84 protocol, we need
a linear code $C_1$ to be efficiently decodeable,
which is used for error correction in step (\ref{error-correction}).
Under the conditions (\ref{con-a})--(\ref{con-c}),
it is difficult to find a pair of linear codes $C_1$ and $C_2$
of which $C_1$ is efficiently decodeable.
On the other hand, since we do not decode $C_2^\bot$ in the BB84 protocol,
we can evaluate the condition (\ref{con-c}) with an arbitrary
decoding method.
Therefore, first we choose a code $C_1$ that satisfies the 
condition (\ref{con-b}) and is efficiently decodeable.
Then we will find a code $C_2$ that 
satisfies the conditions (\ref{con-a}) and (\ref{con-c}).
Given a code $C_1$, choosing a code $C_2$ 
with the condition (\ref{con-a}) is same as
choosing a code $C_2^\bot$ that satisfies
($\mbox{\ref{con-a}}^\prime$) $C_1^\bot \subset C_2^\bot$.

If we fix a rate $R$ lower than
$1-\rom{H}(p_0+p_1)$ and choose a code $C_2^\bot$ 
of rate $R$ at random with the condition ($\mbox{\ref{con-a}}^\prime$),
with high probability the condition (\ref{con-c}) will be 
satisfied \cite[Lemma 4]{mayers01}.
In this section, we will prove that if we fix a rate $R$ lower than
$1- \frac{\rom{H}(p_0)+\rom{H}(p_1)}{2}$ and choose a code $C_2^\bot$
of rate $R$ at random with the condition ($\mbox{\ref{con-a}}^\prime$),
with high probability the condition (\ref{con-c}) will be satisfied.
Some ideas used in the proof are borrowed from \cite{matsumoto02,hamada02}.

We present the main theorem in Section \ref{code-pri},
and the proof of this theorem in Section \ref{proof}.
Then we consider the key rate of securely shared key in Section \ref{achi},
and compare our result with Mayers' in Section \ref{compare}.

\subsection{The code for privacy amplification}
\label{code-pri}
Given a code $C_1^\bot$ of dimension $r$, 
we consider how to choose a code $C_2^\bot$.
Fix a rate $R= \frac{r+m}{n} < 1 -
\frac{\rom{H}(p_0)+\rom{H}(p_1)}{2}$,
and let  
\begin{eqnarray*}
A_m &=& \left\{ C_2^\bot \subset \mathbf{F}_2^n \mid  C_2^\bot \ 
\mbox{is a linear space}, \right. \\ 
&& \hspace{10mm}
\left.  \dim C_2^\bot = r+m,\ C_1^\bot \subset C_2^\bot \right\} 
\label{random-code}
\end{eqnarray*}
be the set from which we choose a code $C_2^\bot$.

\subsubsection{Minimum conditional entropy decoding}
To evaluate the decoding error probability,
we employ the minimum conditional entropy 
decoding \cite[Appendix B]{hamada04}.
Let $\bm{e}$ be an error occurred in $n$ bits binary vector,
$P_{\zeta}$ be the type of first $\frac{n}{2}$ bits of $\bm{e}$,
and $P_{\eta}$ be the type of latter $\frac{n}{2}$ bits of $\bm{e}$.
Then we define the conditional entropy of $\bm{e}$ as
\begin{eqnarray*}
\rom{H}_c(\bm{e}) = \rom{H}_c(P_{\zeta},P_{\eta})
= \frac{\rom{H}(P_{\zeta})+\rom{H}(P_{\eta})}{2}.
\end{eqnarray*}
In the minimum conditional entropy decoding,
we find an estimated error $\hat{\bm{e}}$ that 
minimizes $\rom{H}_c(\hat{\bm{e}})$ and
has the same syndrome, i.e., $\hat{\bm{e}} H_2^T = \bm{e} H_2^T$,
where $H_2$ is parity check matrix of $C_2^\bot$.

\subsubsection{The decoding process of a CSS code}
\label{decode-css}
We assume that only phase errors occur
because we will consider the decoding process by $C_2^\bot$.
Assume that a codeword $\ket{\bm{v}+C_2}$ is sent and
$\sigma_z^{[\bm{e}]} \ket{\bm{v}+C_2}$ is received,
where
\begin{eqnarray*}
\ket{\bm{v}+C_2} &=& \frac{1}{|C_2|^{1/2}}
\sum_{\bm{w} \in C_2}\ket{\bm{v}+\bm{w}} \hspace{3mm} \bm{v} \in C_1, \\
\sigma_z^{[\bm{e}]} &=& \sigma_z^{e_1} \otimes \cdots \otimes \sigma_z^{e_n}, \\
\bm{e} &=& (e_1,\cdots,e_n).
\end{eqnarray*}
Compute the syndrome $\bm{e} H_2^T$ and find an estimated error
$\hat{\bm{e}}$.
Then, apply the unitary operator $\sigma_z^{[\hat{\bm{e}}]}$ to
$\sigma_z^{[\bm{e}]} \ket{\bm{v}+C_2}$ to correct the error.
If 
\begin{eqnarray*}
\sigma_z^{[\bm{e}+\hat{\bm{e}}]} \ket{\bm{v}+C_2}
\neq \ket{\bm{v}+C_2},
\end{eqnarray*}
a decoding error occurs.

\subsubsection{Errors causing decoding errors}
\label{de-error}

We consider when decoding errors occur.
Note that the condition $\hat{\bm{e}} H_2^T = \bm{e} H_2^T$ is
equivalent to
$\bm{e}+\hat{\bm{e}} \in C_2^\bot$.
For a linear code $C_2^\bot$, if there exists a vector
$\hat{\bm{e}}$ such that $\bm{e}+\hat{\bm{e}} \in C_2^\bot$ and
$\rom{H}_c(\hat{\bm{e}}) \le \rom{H}_c(\bm{e})$,
the estimated error is $\hat{\bm{e}}$ instead of $\bm{e}$.
If the unitary operator
$\sigma_z^{[\bm{e}+\hat{\bm{e}}]}$ applied to
a codeword of a CSS code $\ket{\bm{v}+C_2}$, then we have 
\begin{eqnarray*}
&& \sigma_z^{[\bm{e}+\hat{\bm{e}}]} \frac{1}{|C_2|^{1/2}} 
\sum_{\bm{w} \in C_2} \ket{\bm{v}+\bm{w}} \\
&=& \frac{1}{|C_2|^{1/2}} 
\sum_{\bm{w} \in C_2} (-1)^{(\bm{v}+\bm{w})\cdot (\bm{e}+\hat{\bm{e}})} 
\ket{\bm{v}+\bm{w}} .
\end{eqnarray*}
If $(\bm{v}+\bm{w}) \cdot (\bm{e}+\hat{\bm{e}}) = 0$ for all
$\bm{v}+\bm{w} \in C_1$, the codeword is left unchanged by
multiplication of $\sigma_z^{[\bm{e}+\hat{\bm{e}}]}$.
Because $\bm{v} \in C_1$, $\bm{w} \in C_2$, and $C_2 \subset C_1$,
if $\bm{e}+\hat{\bm{e}} \in C_1^\bot$,
then $(\bm{v}+\bm{w}) \cdot (\bm{e}+\hat{\bm{e}}) = 0$.
Thus, if $\bm{e}+\hat{\bm{e}} \in C_1^\bot$,
the errors are not estimated correctly but the received state will be
corrected to the original state, and these errors do not
yield decoding errors.
In case of decoding a CSS code, we define the set of errors
for each $C_2^\bot$,
which cause decoding errors, as 
\begin{eqnarray*}
\label{error-set}
{\cal E}(C_2^\bot) 
= \left\{\bm{e} \in \mathbf{F}_2^n \mid \exists \hat{\bm{e}} \ 
\rom{H}_c(\hat{\bm{e}})\le \rom{H}_c(\bm{e}), \ 
 \bm{e}+\hat{\bm{e}} 
\in C_2^\bot \backslash C_1^\bot \right\}. 
\end{eqnarray*}

\begin{definition}
\label{def1}
We define the decoding error probability of $C_2^\bot$ as
a part of a CSS code over a BSC whose crossover probability
of first $\frac{n}{2}$ bits are $p_1^\prime$ and that of latter
$\frac{n}{2}$ bits are $p_0^\prime$ as
\begin{eqnarray*}
P_{err}(C_2^\bot,p_0^\prime,p_1^\prime) =
\sum_{\bm{e} \in {\cal E}(C_2^\bot)} Q(\bm{e}),
\end{eqnarray*}
where $Q(\bm{e})$ is a probability that $\bm{e}$ occurs in a
BSC whose crossover probability of first $\frac{n}{2}$ bits 
are $p_1^\prime$ and that of latter $\frac{n}{2}$ bits 
are $p_0^\prime$.
\end{definition}
The decoding error probability of $C_1$ as a part of a
CSS code is defined in the same way considering
the decoding process of bit flip errors of a CSS code.
 
\begin{thm}
\label{thm1}
If we choose a code $C_2^\bot$ at random
from $A_m$, for arbitrary $\mu > 0$, we have
\begin{eqnarray*}
\Pr\left\{
P_{err}(C_2^\bot,p_0^\prime,p_1^\prime) \le 
(\frac{n}{2}+1)^4 2^{-n(E(R,p_0,p_1)-\mu)} \right. \\ 
\left. \phantom{\frac{n}{2}}\forall p_0^\prime \le p_0,\ p_1^\prime \le p_1
\right\}
\ge 1 - (\frac{n}{2}+1)^2 2^{-\mu n},              
\end{eqnarray*}
where
\begin{eqnarray*}
&& E(R,p_0,p_1) = \\ 
&& \min_{q_0,q_1}
\left[
\frac{D(q_1|p_1)+D(q_0|p_0)}{2}
+ |1-R-\rom{H}_c(q_1,q_0)|^+
\right],
\end{eqnarray*}
and $|x|^+ = \max\{x,0\}$.
Note that $\min_{q_0,q_1}$ is taken over 
$0 \le q_0,q_1 \le 1$.
Because $\frac{D(q_1|p_1)+D(q_0|p_0)}{2}=0$ only if
$q_1=p_1$, $q_0=p_0$, 
and $R < 1-\frac{\rom{H}(p_0)+\rom{H}(p_1)}{2}$, we have
$E(R,p_0,p_1) > 0$.
\end{thm}

Consequently, we can obtain a code $C_2^\bot$ that satisfy the 
condition (\ref{con-c}) with high probability
by choosing a code at random from $A_m$.

\subsection{Proof of the theorem}
\label{proof}

Refer to \cite{csiszar-korner81} for the method of type
used in this section.
We also use the notation
$T_{(P_{\zeta},P_{\eta})}^n$ as the set of binary vectors whose type of 
first $\frac{n}{2}$ bits is $P_{\zeta}$ and that of 
latter $\frac{n}{2}$ bits is $P_{\eta}$.
$P_{\frac{n}{2}}^2$ is the direct product of the sets of all possible
types over $\{0,1\}^{\frac{n}{2}}$, i.e., $P_{\frac{n}{2}} \times
P_{\frac{n}{2}}$.

We classify ${\cal E}(C_2^\bot)$ by the types in $P_{\frac{n}{2}}^2$ as
\begin{eqnarray*}
{\cal E}(C_2^\bot) = \bigcup_{(P_\zeta,P_\eta)} 
{\cal E}_{(P_\zeta,P_\eta)}(C_2^\bot),
\end{eqnarray*}
where 
${\cal E}_{(P_\zeta,P_\eta)}(C_2^\bot)= {\cal E}(C_2^\bot) \cap T_{(P_\zeta,P_\eta)}^n$.
First, we prove that if we choose a code $C_2^\bot$ at random
from $A_m$, $C_2^\bot$ satisfies the following property
with high probability. Then we prove that 
the decoding error probability of 
$C_2^\bot$
that satisfy the following property is small.
Given arbitrary $\mu > 0$,
for all types $(P_\zeta,P_\eta) \in P_{\frac{n}{2}}^2$,
\begin{eqnarray*}
\label{prop-uni}
\frac{|{\cal E}_{(P_\zeta,P_\eta)}(C_2^\bot)|}{|T_{(P_\zeta,P_\eta)}^n|}
\le 2^{-n(|1-\rom{H}_c(P_\zeta,P_\eta)-R|^+ - \mu)}.
\end{eqnarray*}
To prove this, we evaluate the average of 
$\frac{|{\cal
E}_{(P_\zeta,P_\eta)}(C_2^\bot)|}{|T_{(P_\zeta,P_\eta)}^n|}$
over $C_2^\bot \in A_m$.
Define the set of codes that cannot correct $\bm{e}$ as 
\begin{eqnarray*}
B_m(\bm{e}) = \left\{
C_2^\bot \in A_m \mid \bm{e} \in {\cal E}(C_2^\bot)
\right\}.
\end{eqnarray*}
Define $C_m(\bm{e})$ as
\begin{eqnarray*}
C_m(\bm{e}) = \left\{ C_2^\bot \in A_m \mid \bm{e} \in C_2^\bot \backslash C_1^\bot \right\} 
\end{eqnarray*}
and $G$ as the set of bijective linear maps $\alpha$ on $\mathbf{F}_2^n$
that satisfies $\alpha (C_1^\bot) = C_1^\bot$.
Then we have the following equalities:
\begin{eqnarray*}
&& |C_m(\bm{e})| \\
&=& \left|\left\{ C_2^\bot \in A_m \mid \bm{e} \in C_2^\bot 
\backslash C_1^\bot \right\} \right|  \\
&=& \left| \left\{ \alpha (C_2^\bot) \mid \bm{e} \in \alpha (C_2^\bot 
\backslash C_1^\bot),\alpha \in G , \
\mbox{$C_2^\bot$ is fixed}\right\} \right| \\ 
&=& \left| \left\{ \beta\alpha (C_2^\bot) 
\mid \beta (\bm{e}) \in \beta\alpha (C_2^\bot 
\backslash C_1^\bot), \ 
\alpha,\beta \in G, \right. \right.  \\ 
&& \left. \left.  \hspace{10mm}
\mbox{$\beta$ and $C_2^\bot$ are fixed} \right\} \right|. 
\end{eqnarray*}
Since there exists $\beta \in G$ such that $\bm{e}^\prime = \beta(\bm{e})$
for arbitrary 
$\bm{e}$ and $\bm{e}^\prime \in \mathbf{F}_2^n \backslash C_1^\bot$,
$|C_m(\bm{e})|$ does not depend on
$\bm{e} \in \mathbf{F}_2^n \backslash C_1^\bot$ and
\begin{eqnarray*}
|C_m(\bm{e})| &=& 
\frac{\sum_{\bm{e} \in \mathbf{F}_2^n \backslash C_1^\bot}
|C_m(\bm{e})|}{|\mathbf{F}_2^n \backslash C_1^\bot|} \\
&=& \frac{\sum_{\bm{e} \in \mathbf{F}_2^n \backslash C_1^\bot}
|\{C_2^\bot \in A_m \mid \bm{e} \in C_2^\bot \backslash C_1^\bot \}|}
{|\mathbf{F}_2^n \backslash C_1^\bot|} \\
&=& \frac{\sum_{C_2^\bot \in A_m} 
|\{\bm{e} \in \mathbf{F}_2^n \backslash C_1^\bot \mid
 \bm{e} \in C_2^\bot \backslash C_1^\bot\}|}{|\mathbf{F}_2^n \backslash C_1^\bot |} \\
&=& \frac{|C_2^\bot \backslash C_1^\bot||A_m|}{|\mathbf{F}_2^n 
\backslash C_1^\bot|}. 
\end{eqnarray*}
From the definition, it is obvious that $|C_m(\bm{e})|=0$
for $\bm{e} \in C_1^\bot$.
Hence
\begin{eqnarray*}
|C_m(\bm{e})| &\le& 
\frac{|C_2^\bot \backslash C_1^\bot||A_m|}{|\mathbf{F}_2^n 
\backslash C_1^\bot|} \\*
&=& \frac{2^{r+m}-2^r}{2^n - 2^r}|A_m| \\*
&=& \frac{|A_m|}{2^{n-(r+m)}}\frac{1-2^{-m}}{1-2^{-n+r}} \\*
&\le& \frac{|A_m|}{2^{n-(r+m)}} \\*
&=& |A_m| 2^{-n(1-R)}.
\end{eqnarray*}

Because the condition for $C_2^\bot \in A_m$ to belong to $B_m(\bm{e})$ is 
$\exists \hat{\bm{e}} \ 
\rom{H}_c(\hat{\bm{e}}) \le 
\rom{H}_c(\bm{e}),\bm{e}+\hat{\bm{e}} \in C_2^\bot \backslash
C_1^\bot$, we obtain
\begin{eqnarray*}
\frac{|B_m(e)|}{|A_m|}  &\le&
\frac{1}{|A_m|}
\sum_{\scriptstyle \hat{\bm{e}} \in \mathbf{F}_2^n \atop 
\rom{H}_c(\hat{\bm{e}}) \le \rom{H}_c(\bm{e})} 
|C_m(\bm{e}+\hat{\bm{e}})| \\ 
&\le& \sum_{\scriptstyle \hat{\bm{e}} \in \mathbf{F}_2^n \atop 
\rom{H}_c(\hat{\bm{e}})  \le \rom{H}_c(\bm{e})} 2^{-n(1-R)},
\end{eqnarray*} 
while $\frac{|B_m(e)|}{|A_m|} \le 1$.
Let $|x|^+=\max\{x,0\}$ and 
note that if $a,b \ge 0$, then 
$\min\{a+b,1\} \le \min\{a,1\}+\min\{b,1\}$.

Using above definitions,
we have
\begin{eqnarray*}
&& \frac{1}{|A_m|} \sum_{C_2^\bot \in A_m}
\frac{|{\cal E}_{(P_\zeta,P_\eta)}(C_2^\bot)|}{|T_{(P_\zeta,P_\eta)}^n|} \\
&=& \frac{1}{|T_{(P_\zeta,P_\eta)}^n|} 
\sum_{\bm{e} \in T_{(P_\zeta,P_\eta)}^n}
\frac{|B_m(\bm{e})|}{|A_m|}  \\
&\le& \frac{1}{|T_{(P_\zeta,P_\eta)}^n|}
\sum_{\bm{e} \in T_{(P_\zeta,P_\eta)}^n}
\min\left\{ \sum_{\scriptstyle \hat{\bm{e}} \in \mathbf{F}_2^n \atop 
\rom{H}_c(\hat{\bm{e}})  \le \rom{H}_c(\bm{e})} 
2^{-n(1-R)},1\right\}  \\
&=& \min \left\{ \sum_{\scriptstyle (P_{\zeta}^\prime,P_{\eta}^\prime) 
\in P_{\frac{n}{2}}^2 \atop
\rom{H}_c(P_{\zeta}^\prime,P_{\eta}^\prime) 
\le \rom{H}_c(P_{\zeta},P_{\eta})}
|T_{(P_{\zeta}^\prime,P_{\eta}^\prime)}^n| 2^{-n(1-R)} ,1 \right\} \\
&\le& \sum_{\scriptstyle (P_{\zeta}^\prime,P_{\eta}^\prime) 
\in P_{\frac{n}{2}}^2 \atop
\rom{H}_c(P_{\zeta}^\prime,P_{\eta}^\prime) 
\le \rom{H}_c(P_{\zeta},P_{\eta})}
2^{-n| 1-R-\rom{H}_c(P_{\zeta}^\prime,P_{\eta}^\prime)|^+} \\
&\le& |P_{\frac{n}{2}}^2|
\max_{\scriptstyle (P_{\zeta}^\prime,P_{\eta}^\prime) 
\in P_{\frac{n}{2}}^2 \atop
\rom{H}_c(P_{\zeta}^\prime,P_{\eta}^\prime) 
\le \rom{H}_c(P_{\zeta},P_{\eta}) }
2^{-n|1-R-\rom{H}_c(P_{\zeta}^\prime,P_{\eta}^\prime)|^+ } \\
&\le& (\frac{n}{2}+1)^2
2^{-n|1-R-\rom{H}_c(P_{\zeta},P_{\eta})|^+ }
\end{eqnarray*}

Let $A_b(\mu,P_\zeta,P_\eta)$ and 
$A_g(\mu)$ be
\begin{eqnarray*}
&& A_b(\mu,P_\zeta,P_\eta)
= \left\{
C_2^\bot \in A_m \ \Bigg| \
\frac{|{\cal E}_{(P_\zeta,P_\eta)}(C_2^\bot)|}{|T_{(P_\zeta,P_\eta)}^n|} 
\right. \\
&& \left.  
\phantom{\frac{|{\cal E}_{(P_\zeta,P_\eta)}(C_2^\bot)|}{|T_{(P_\zeta,P_\eta)}^n|}}
> (\frac{n}{2}+1)^2
2^{-n(|1-R-\rom{H}_c(P_{\zeta},P_{\eta})|^+ - \mu)}
\right\}, \\
&& A_g(\mu) = 
A_m \backslash \bigcup_{(P_\zeta,P_\eta) \in P_{\frac{n}{2}}^2}
A_b(\mu,P_\zeta,P_\eta).
\end{eqnarray*}
From the union bound and the Chebychev inequality,
we have
\begin{eqnarray*}
&& \frac{|A_g(\mu)|}{|A_m|} \\*
&=& 1 - \frac{\bigcup_{(P_\zeta,P_\eta) \in P_{\frac{n}{2}}^2}
|A_g(\mu,P_\zeta,P_\eta)|}{|A_m|} \\*
&\ge& 1 - \sum_{(P_\zeta,P_\eta) \in P_{\frac{n}{2}}^2}
\frac{|A_g(\mu,P_\zeta,P_\eta)|}{|A_m|} \\*
&\ge& 1 - \sum_{(P_\zeta,P_\eta) \in P_{\frac{n}{2}}^2}
\frac{(\frac{n}{2}+1)^2
2^{-n|1-R-\rom{H}_c(P_{\zeta},P_{\eta})|^+}}
{(\frac{n}{2}+1)^2
2^{-n(|1-R-\rom{H}_c(P_{\zeta},P_{\eta})|^+ - \mu)}} \\*
&\ge& 1 - (\frac{n}{2}+1)^2 2^{-\mu n}
\end{eqnarray*}

Next, we evaluate the decoding error probability of $C_2^\bot \in A_g(\mu)$.
Let $p_0^\prime \le p_0$ and $p_1^\prime \le p_1$,
and $Q(\bm{e})$ be a probability that $\bm{e}$ occurs
in a BSC whose crossover probability of first $\frac{n}{2}$ bits
are $p_1^\prime$ and that of latter $\frac{n}{2}$ bits are
$p_0^\prime$.
Then the decoding error probability of $C_2^\bot$ as a part of
a CSS code is
\begin{eqnarray*}
&& P_{err}(C_2^\bot,p_0^\prime,p_1^\prime) \\
&=& \sum_{\bm{e} \in {\cal E}(C_2^\bot)} Q(\bm{e}) \\
&=& \sum_{(P_\zeta,P_\eta) \in P_{\frac{n}{2}}^2}
\sum_{\bm{e} \in {\cal E}_{(P_\zeta,P_\eta)}(C_2^\bot)}
Q(\bm{e}) \\
&=& \sum_{(P_\zeta,P_\eta) \in P_{\frac{n}{2}}^2}
\frac{|{\cal E}_{(P_\zeta,P_\eta)}(C_2^\bot)|}{|T_{(P_\zeta,P_\eta)}^n|}
Q(T_{(P_\zeta,P_\eta)}^n) \\
&\le& \sum_{(P_\zeta,P_\eta) \in P_{\frac{n}{2}}^2}
(\frac{n}{2}+1)^2
2^{-n(|1-R-\rom{H}_c(P_{\zeta},P_{\eta})|^+ - \mu)} \\
&& \times 2^{-\frac{n}{2}\{
D(P_\zeta(1)|p_1^\prime)+D(P_\eta(1)|p_0^\prime)
\}} \\
&\le& (\frac{n}{2}+1)^4 2^{-n(E(R,p_0^\prime,p_1^\prime)-\mu)},
\end{eqnarray*}
where 
\begin{eqnarray*}
&& E(R,p_0^\prime,p_1^\prime) = \\
&& \min_{q_0,q_1}
\left[
\frac{D(q_1|p_1^\prime)+D(q_0|p_0^\prime)}{2}
+ |1-R-\rom{H}_c(q_1,q_0)|^+
\right].
\end{eqnarray*}
\begin{lem}
\label{lemma-E}
\begin{eqnarray*}
\min_{\scriptsize 0 \le p_0^\prime \le p_0 \atop
0 \le p_1^\prime \le p_1}
E(R,p_0^\prime,p_1^\prime)
= E(R,p_0,p_1).
\end{eqnarray*}
\end{lem}
We prove this lemma in Appendix \ref{app2}.
From this lemma, we have
\begin{eqnarray*}
&& P_{err}(C_2^\bot,p_0^\prime,p_1^\prime) \\
&\le& (\frac{n}{2}+1)^4 2^{-n(E(R,p_0,p_1)-\mu)} \ \ 
\forall p_0^\prime \le p_0,\ p_1^\prime \le p_1.
\end{eqnarray*}
Then Theorem~\ref{thm1} is proved.

\subsection{Achievable key rate}
\label{achi}
We proved that if we fix a code $C_1^\bot$ and choose a code $C_2^\bot$ of
a fixed rate $R<1-\frac{\rom{H}(p_0)+\rom{H}(p_1)}{2}$ at random
with the condition $C_1^\bot \subset C_2^\bot$,
the decoding error probability of $C_2^\bot$
as a part of a CSS code is small with high probability.
Consequently, we can conduct random privacy amplification
with a code $C_2$ of a rate 
higher than $\frac{\rom{H}(p_0)+\rom{H}(p_1)}{2}$.

If we estimate an error rate in a lump
(test bits in each basis are lumped together and
a single error rate is computed),
an estimated error rate is $\frac{p_0+p_1}{2}$ instead of
$p_0$ and $p_1$.
Thus, we can conduct random privacy amplification
with a code $C_2$ with a rate 
higher than $\rom{H}\left(\frac{p_0+p_1}{2}\right)$.
Since the entropy function is concave, 
we can conduct random privacy amplification with a code $C_2$ of 
a lower rate by estimating error rates 
separately, which enables us to share a longer key. 

\subsection{Comparison with Mayers' evaluation}
\label{compare}
In this section, we compare our result with
Mayers'.
In case of estimating an error rate in a lump,
we can conduct random privacy amplification with 
a code $C_2$ of a rate higher than $\rom{H}\left(\frac{p_0+p_1}{2}\right)$.
Because there exists efficiently decodeable codes 
whose rate is fairly close to $1-\rom{H}\left(\frac{p_0+p_1}{2}\right)$
and whose decoding error probability is small,
we can securely share a key with a rate lower than 
$1-2\rom{H}\left(\frac{p_0+p_1}{2}\right)$.
With Mayers' evaluation of minimum Hamming weight of
$C_2^\bot \backslash C_1^\bot$ in \cite{mayers01},
we can securely share a key with a rate lower than 
$1-\rom{H}\left(\frac{p_0+p_1}{2}\right)-\rom{H}(p_0+p_1)$,
where $1-\rom{H}\left(\frac{p_0+p_1}{2}\right)$ is the rate of $C_1$
for error correction and $\rom{H}(p_0+p_1)$ is the rate of
$C_2$ for privacy amplification.

\begin{figure}[t!]
\includegraphics*[width=\linewidth]{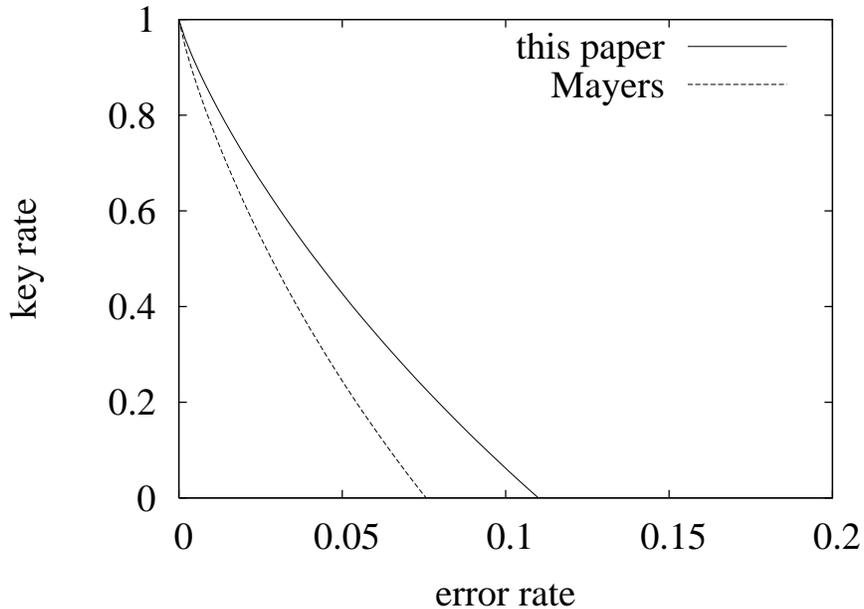}
      \caption{Comparison of achievable key rates.}
      \label{error-key-rate}
\end{figure}

According to the evaluation of the decoding error probability,
we showed that the tolerable error rate can be increased
from 7.5 \% to 11 \% in the BB84 protocol with
random privacy amplification. 
Figure \ref{error-key-rate} shows  
the secure key rate 
of this paper and Mayers' against error rate,
and the key rate falls to $0$ at the point of 7.5 \% and 11 \%
respectively.

\section{Conclusion}
\label{conclusion}

For a fixed code $C_1$,
we showed that we can decrease a rate of randomly chosen code $C_2$,
keeping the BB84 protocol to be secure.
Consequently, we proved that
the BB84 protocol with random privacy amplification
can tolerate severer noise and can share longer keys.


\section*{Acknowledgment}
We appreciate the helpful comment by Dr.~Masahito Hayashi 
on an earlier version of this paper.
This research is partly supported by 
the Japan Society for the Promotion of Science.

\appendix
\section{Proof of theorem \ref{theorem-security}}
\label{appendix-security}

In this appendix,
we prove Theorem \ref{theorem-security}, i.e.,
we prove that we can securely share a key 
against general eavesdropping attacks with 
the linear codes $C_1$ and $C_2$ that satisfy
the conditions (\ref{con-a})--(\ref{con-c}) in 
Section \ref{security-condition}.
Basic ideas in this proof are borrowed from \cite{hamada04,gottesman03}.

The outline of this proof is similar to the
proof in \cite{hamada04}.
We introduce a class of CSS codes and the CSS code protocol 
in Sections \ref{a-class-of-css-codes} and 
\ref{css-protocol} respectively.
Then we define some notations used in this proof in Section \ref{notation}.
After that, we define Eve's attack mathematically and 
relate the BB84 protocol to the CSS code protocol in 
Sections \ref{eve's-attack} and 
\ref{relating-to-css-protocol} respectively.
In Section \ref{bounding-the-fidelity},
in order to bound the mutual information,
we bound the fidelity, which is defined in
Section \ref{relating-to-css-protocol}.
Finally, we upper bound the mutual information
of a shared key and Eve's accessible information 
in Section \ref{bounding-the-mutual-information} using 
the result in Section \ref{bounding-the-fidelity}.


\subsection{A class of CSS codes}
\label{a-class-of-css-codes}

In this section, we define a class of CSS codes.
A CSS code ${\cal Q}$ is constructed from two linear codes $C_1$ and 
$C_2$ that satisfy $C_2 \subset C_1$.
A codeword $\ket{\phi_\bm{u}} \in {\cal Q}$ is
\begin{eqnarray*}
\ket{\phi_{\bm{u}}} = 
\frac{1}{\sqrt{|C_2|}} \sum_{\bm{w} \in C_2} 
\ket{\bm{u}+\bm{w}},
\end{eqnarray*}
where $\bm{u}$ is a coset representative of $C_1 \slash C_2$.
A class of CSS codes is $\{{\cal Q}_{\bm{x}\bm{z}}\}$ that are 
parametrized by coset representatives 
$\bm{x} \in \mathbf{F}_2^n \slash C_1$ and 
$\bm{z} \in \mathbf{F}_2^n \slash C_2^\bot$,
and a codeword $\ket{\phi_{\bm{u}\bm{x}\bm{z}}} \in {\cal Q}_{\bm{x}\bm{z}}$
is
\begin{eqnarray*}
\ket{\phi_{\bm{u}\bm{x}\bm{z}}} = 
\frac{1}{\sqrt{|C_2|}} \sum_{\bm{w} \in C_2} 
(-1)^{\bm{z}\cdot\bm{w}} \ket{\bm{u}+\bm{w}+\bm{x}}. 
\end{eqnarray*}
A recovery operator for a CSS code ${\cal Q}_{\bm{x}\bm{z}}$
is a TPCP (Trace Preserving Completely Positive) map
on $\rom{L}({\cal H}^{\otimes n})$ that represent
the measurement of syndrome and the unitary operation
of error correction,
where ${\cal H}$ is a Hilbert space of dimension $2$,
and $\rom{L}({\cal H})$ is the linear space of operators
on ${\cal H}$.
We denote by ${\cal R}_{\bm{x}\bm{z}}$ the recovery
operator for ${\cal Q}_{\bm{x}\bm{z}}$.


\subsection{The CSS code protocol}
\label{css-protocol}

To prove the security of the BB84 protocol, we relate the BB84 protocol with
the following CSS code protocol \cite[Section 4]{hamada04}.
In this section, we use bold large letters for random 
variables and small bold letters for their realizations,
e.g., $\bm{x}$ denote a realization of a random variable
$\bm{X}$.
We also use a notation $\mathbb{E}_{\bm{x}}$ as
the expectation operator over $\bm{X}$.

Suppose Alice chooses  coset representatives $\bm{u} \in C_1\slash C_2$,
$\bm{x} \in \mathbf{F}_2^n \slash C_1$,
$\bm{z} \in \mathbf{F}_2^n \slash C_2^\bot$ at random
and send $\ket{\phi_{\bm{u}\bm{x}\bm{z}}}$.
Let ${\cal A}_n$ be a TPCP map on
$\rom{L}({\cal H}^{\otimes n})$ representing Eve's eavesdropping attack,
and $\bm{E}$ be a random variable that is Eve's measurement result.
According to \cite[Section 5.3]{schumacher96},
we can bound the mutual information between $\bm{U}$ and $\bm{E}$ as
\begin{eqnarray}
\rom{I}(\bm{U};\bm{E}|\bm{X}=\bm{x},\bm{Z}=\bm{z}) \le 
S_{\bm{x}\bm{z}},
\label{mutual}
\end{eqnarray}
where $S_{\bm{x}\bm{z}}$ is the entropy exchange \cite[Section 5]{schumacher96}
after the system suffers a Eve's attack ${\cal A}_n$ and recovery operator
${\cal R}_{\bm{x}\bm{z}}$, i.e.,
\begin{eqnarray*}
S_{\bm{x}\bm{z}} &=& 
S(\rho), \\
\rho &=& 
[{\cal I}_n\otimes{\cal R}_{\bm{x}\bm{z}}](
[{\cal I}_n\otimes{\cal A}_n](
\ket{\Phi_{\bm{x}\bm{z}}}\bra{\Phi_{\bm{x}\bm{z}}})), \\
\ket{\Phi_{\bm{x}\bm{z}}} &=&
\frac{1}{\sqrt{|C_1\slash C_2|}} \sum_{\bm{u} \in C_1 \slash C_2}
\ket{\phi_{\bm{u}\bm{x}\bm{z}}} \otimes \ket{\phi_{\bm{u}\bm{x}\bm{z}}},
\end{eqnarray*}
and $S(\cdot)$ denotes the von Neumann entropy.
Let $F_{\bm{x}\bm{z}}$ be the entanglement fidelity 
of above process, i.e.,
\begin{eqnarray*}
F_{\bm{x}\bm{z}} = \bra{\Phi_{\bm{x}\bm{z}}}\rho\ket{\Phi_{\bm{x}\bm{z}}}.
\end{eqnarray*}
Then, by the quantum Fano inequality,
Eq.~(24) of  \cite[Section 6.2]{schumacher96},
$S_{\bm{x}\bm{z}}$ is bounded as
\begin{eqnarray}
\label{fano}
S_{\bm{x}\bm{z}} \le \rom{H}(1-F_{\bm{x}\bm{z}}) + (1-F_{\bm{x}\bm{z}})2n,
\end{eqnarray}
where $\rom{H}(\cdot)$ is the binary entropy function.
Combining Eqs.~(\ref{mutual}) and (\ref{fano}) and taking the average of
the both sides over $\bm{x},\bm{z}$, we have
\begin{eqnarray}
\rom{I}(\bm{U};\bm{E}|\bm{X},\bm{Z}) \le
\rom{H}(1-\mathbb{E}_{\bm{x}\bm{z}}F_{\bm{x}\bm{z}}) 
+ (1-\mathbb{E}_{\bm{x}\bm{z}}F_{\bm{x}\bm{z}})2n,
\label{eve-mutual}
\end{eqnarray}
where we used concavity of entropy function, i.e.,
$\mathbb{E}_{\bm{x}\bm{z}}\rom{H}(1-F_{\bm{x}\bm{z}}) \le \rom{H}(1-\mathbb{E}_{\bm{x}\bm{z}}F_{\bm{x}\bm{z}})$.
From Eq.~(27) of \cite[Section 5]{hamada04},
the entanglement fidelity $F_{\bm{x}\bm{z}}$ is bounded as
\begin{eqnarray}
\label{fid}
1-\mathbb{E}_{\bm{x}\bm{z}}F_{\bm{x}\bm{z}} \le \sum_{(\bm{e}_x,\bm{e}_z) \in {\cal E}}
{\cal P}(\bm{e}_x,\bm{e}_z),
\end{eqnarray}
where ${\cal E}$ is the set of uncorrectable errors of a CSS code ${\cal Q}$,
\begin{eqnarray*}
{\cal P}(\bm{e}_x,\bm{e}_z) &=&
\bra{\Psi_{\bm{e}_x\bm{e}_z}^n}
[{\cal I}_n \otimes {\cal A}_n](\ket{\Psi^n}\bra{\Psi^n})
\ket{\Psi_{\bm{e}_x\bm{e}_z}^n}, \\
\ket{\Psi^n} &=& \frac{1}{\sqrt{2^n}}
\sum_{\bm{l} \in \mathbf{F}_2^n}
\ket{\bm{l}} \otimes \ket{\bm{l}}, \\
\ket{\Psi_{\bm{e}_x\bm{e}_z}^n} &=& \frac{1}{\sqrt{2^n}}
\sum_{\bm{l} \in \mathbf{F}_2^n}
\ket{\bm{l}} \otimes \sigma_x^{[\bm{e}_x]}\sigma_z^{[\bm{e}_z]}\ket{\bm{l}}.
\end{eqnarray*}


\subsection{Notation}
\label{notation}

First, we fix the positions of remaining $2n$ bits
out of $(4+\theta)n$ bits in step (\ref{divide}),
and represent these positions by $T$.
Lets $\bm{a}_T,\bm{b}_T \in \mathbf{F}_2^{2n}$ be
the subsequences of $\bm{a},\bm{b}$ that correspond to $T$, which
include $n$ bits where $a_i=b_i=0$ and 
$n$ bits where $a_i=b_i=1$. 
We further divide the positions $T$ into four blocks,
$T_0^k,T_0^t,T_1^k,T_1^t$.
$T_0^k,T_0^t$ consist of  the positions
that $a_i=b_i=0$, and $T_1^k,T_1^t$ consist of the 
positions that $a_i=b_i=1$.
$T_0^k,T_1^k$ are the positions used for generating a key, and
$T_0^t,T_1^t$ are the positions used for estimating an error rate.
$T_0^k,T_0^t,T_1^k,T_1^t$ depend on $\bm{a}_T,\bm{b}_T,\bm{pos}$,
where $\bm{pos}$ represents
how to divide the remaining positions $T$
into the positions $T^k$ for generating a key and the positions $T^t$
for estimating an error rate. Note that $T^k$ consists of $T_0^k$ and
$T_1^k$, and $T^t$ consists of $T_0^t$ and $T_1^t$. 

We also use a notation $\bm{k}_T$ as 
a subsequence of $\bm{k} \in \mathbf{F}_2^{(4+\theta)n}$ 
that corresponds to $T$, and $\bm{k}_{T_0^k}$, $\bm{k}_{T_1^k}$,
$\bm{k}_{T_0^t}$, $\bm{k}_{T_1^t}$ $\bm{k}_{T^k}$,
$\bm{k}_{T^t}$are subsequences of $\bm{k}_T$ 
that corresponds to $T_0^k$, $T_1^k$, $T_0^t$, $T_1^t$, $T^k$, $T^t$
respectively. Subsequences of 
$\bm{c},\bm{d},\tilde{\bm{k}} \in \mathbf{F}_2^{(4+\theta)n}$ are
defined in the same way.  


\subsection{Eve's attack}
\label{eve's-attack}

Let a TPCP map 
${\cal A}: \rom{L}({\cal H}^{\otimes (4+\theta)n}) \to \rom{L}({\cal H}^{\otimes (4+\theta)n})$ 
represent Eve's eavesdropping attack (plus channel noise) 
on transmitted $(4+\theta)n$
qubits.
Note that ${\cal A}$ does not depend on $\bm{a},\bm{b},\bm{pos},\pi$.
In the BB84 protocol, Alice chooses $(4+\theta)n$-bit string $\bm{k}$ and 
sends it with either $\{\ket{0},\ket{1}\}$ or
$\{\ket{+},\ket{-}\}$ basis according to $\bm{a}$, i.e.,
Alice sends $H^{[\bm{a}]} \ket{\bm{k}}$,
where $H^{[\bm{a}]} = H^{a_1} \otimes \cdots \otimes
H^{a_{(4+\theta)}}$,
and $H$ is a Hadamard transformation.
Bob receives 
${\cal A}(H^{[\bm{a}]} \ket{\bm{k}}\bra{\bm{k}}H^{[\bm{a}]})$, and 
measures it by either $\{\ket{0},\ket{1}\}$ or
$\{\ket{+},\ket{-}\}$ basis according to $\bm{b}$ and
obtain $\tilde{\bm{k}}$.
Note that $[H \otimes H] \ket{\Psi} = \ket{\Psi}$, where
\begin{eqnarray*}
\ket{\Psi} = \frac{\ket{0}\ket{0}+\ket{1}\ket{1}}{\sqrt{2}},
\end{eqnarray*}
and that we can denote Eve's attack by a unitary operator on
Bob's system and Eve's system 
${\cal H}^{\otimes (4+\theta)n} \otimes {\cal H}_E$, i.e.,
\begin{eqnarray*}
U_{BE} [ H^{[\bm{a}]} \ket{\bm{k}} \otimes \ket{e_0} ],
\end{eqnarray*}
where $\ket{e}_0$ is a state in ${\cal H}_E$,
and $U_{BE}$ is an unitary operation on 
${\cal H}^{(4+\theta)n} \otimes {\cal H}_E$.
We can mathematically regard Alice's sent bits $\bm{k}$ and 
Bob's received bits $\tilde{\bm{k}}$ as follows.
First, Alice and Bob share a bipartite state
\begin{eqnarray*}
\ket{\Psi^{(4+\theta)n}} &=& 
\frac{1}{\sqrt{2^{(4+\theta)n}}} \sum_{\bm{l} \in \mathbf{F}_2^{(4+\theta)n}}
H^{[\bm{a}]} \ket{\bm{l}} \otimes H^{[\bm{a}]} \ket{\bm{l}} \\
&=& \frac{1}{\sqrt{2^{(4+\theta)n}}} \sum_{\bm{l} \in \mathbf{F}_2^{(4+\theta)n}}
\ket{\bm{l}} \otimes \ket{\bm{l}}.
\end{eqnarray*}
Then Bob's system suffers Eve's attack and the bipartite state becomes
\begin{eqnarray}
\label{entangled}
\ket{\varphi_{ABE}} &=& 
(I_A \otimes U_{BE}) [ \ket{\Psi^{(4+\theta)n}} \otimes \ket{e_0} ] \nonumber \\ 
&=&  \frac{1}{\sqrt{2^{(4+\theta)n}}} \sum_{\bm{l} \in \mathbf{F}_2^{(4+\theta)n}}
H^{[\bm{a}]} \ket{\bm{l}} 
\otimes U_{BE} (H^{[\bm{a}]} \ket{\bm{l}} \otimes \ket{e_0}),
\end{eqnarray} 
where 
$Tr_E[\ket{\varphi_{ABE}}\bra{\varphi_{ABE}}]=[{\cal I} \otimes {\cal
A}](\ket{\Psi^{(4+\theta)n}}\bra{\Psi^{(4+\theta)n}})$,
and $Tr_E[\cdot]$ denotes the partial trace over ${\cal H}_E$.
After that, Alice measures her system by the $\{\ket{0},\ket{1}\}$ basis
or the $\{\ket{+},\ket{-}\}$ basis according to $\bm{a}$, and obtain 
$\bm{k} \in \mathbf{F}_2^{(4+\theta)n}$.
This measurement changes
the state in Eq.~(\ref{entangled}) to
\begin{eqnarray}
\label{projected}
H^{[\bm{a}]} \ket{\bm{k}} \otimes 
U_{BE} [ H^{[\bm{a}]} \ket{\bm{k}} \otimes \ket{e_0} ].
\end{eqnarray}
Then Bob measures his system in Eq.~(\ref{projected})
by the $\{\ket{0},\ket{1}\}$ basis or
the $\{\ket{+},\ket{-}\}$ basis 
according to $\bm{b}$, and obtain $\tilde{\bm{k}}$.

Let 
\begin{eqnarray*}
{\cal P}(\bm{c},\bm{d}) = 
\bra{\Psi_{\bm{c}\bm{d}}^{(4+\theta)n}}
[{\cal I} \otimes {\cal A}](\ket{\Psi^{(4+\theta)n}}\bra{\Psi^{(4+\theta)n}})
\ket{\Psi_{\bm{c}\bm{d}}^{(4+\theta)n}},
\end{eqnarray*}
where $\bm{c},\bm{d} \in \mathbf{F}_2^{(4+\theta)n}$.
Since Alice and Bob measure the qubits by $\{\ket{0},\ket{1}\}$ basis
when $a_i=b_i=0$ and by $\{\ket{+},\ket{-}\}$ basis when $a_i=b_i=1$,
we can relate $\bm{c},\bm{d}$ with $\bm{k},\tilde{\bm{k}}$
as 
\begin{eqnarray*}
\bm{c}_{T_0^t} = \bm{k}_{T_0^t}-\tilde{\bm{k}}_{T_0^t} = \bm{f}_0 \\
\bm{d}_{T_1^t} = \bm{k}_{T_1^t}-\tilde{\bm{k}}_{T_1^t} = \bm{f}_1,
\end{eqnarray*}
where $\bm{f}_0$ and $\bm{f}_1$ are the $\frac{n}{2}$-bit strings
from which we estimate error rates.


\subsection{Relating the BB84 protocol to the CSS code protocol}
\label{relating-to-css-protocol}

In the BB84 protocol, we select the linear codes $C_1$
and $C_2$ that satisfy
the conditions (\ref{con-a})--(\ref{con-c})
from $\bm{f}_0,\bm{f}_1$.
We prove that the following protocol is secure.

Alice randomly selects coset representatives  
$\bm{u},\bm{x},\bm{z}$,
where $\bm{u} \in \pi(C_1)\slash \pi(C_2)$,
$\bm{x} \in \mathbf{F}_2^n \slash \pi(C_1)$,
$\bm{z} \in \mathbf{F}_2^n \slash \pi(C_2^\bot)$, and $\bm{u}$
corresponds to a shared key in the BB84 protocol.
Then Alice selects $\ket{\bm{l}} \otimes \ket{\bm{m}}$,
where $\bm{l} \in \mathbf{F}_2^n$, $\bm{m} \in \mathbf{F}_2^{(2+\theta)n}$, 
$\bm{l}$ corresponds to test bits, and 
$\bm{m}$ corresponds to discarded bits in the BB84 protocol.
Then Alice sends 
\begin{eqnarray*}
H^{[\bm{a}]}[
\ket{\phi_{\bm{u}\bm{x}\bm{z}}} \otimes 
\ket{\bm{l}} \otimes \ket{\bm{m}}]
\end{eqnarray*}
to Bob, where
$H^{[\bm{a}]}= H^{a_1}\otimes \cdots \otimes H^{a_{(4+\theta)n}}$.
Bob measures the test qubits,
and Alice and Bob obtain  $\bm{f}_0,\bm{f}_1$.
Then, Bob corrects errors and obtain a key.

We consider this procedure as follows.
First, Alice and Bob share a bipartite state
\begin{eqnarray*}
\rho =
\ket{\Phi_{\bm{x}\bm{z}}^\prime}\bra{\Phi_{\bm{x}\bm{z}}^\prime}
\otimes \ket{\Psi^n}\bra{\Psi^n} \otimes 
\ket{\Psi^{(2+\theta)n}}\bra{\Psi^{(2+\theta)n}},
\end{eqnarray*}
where
\begin{eqnarray*}
\ket{\Phi_{\bm{x}\bm{z}}^\prime} =
[H^{[\bm{a}_{T^k}]} \otimes H^{[\bm{a}_{T^k}]}]
\ket{\Phi_{\bm{x}\bm{z}}}.
\end{eqnarray*}
Then, Bob's system suffers a Eve's attack ${\cal A}$, and $\rho$ becomes
$\rho^\prime$.
After Alice and Bob obtain measurement disagreements $\bm{f}_0$, $\bm{f}_1$,
bipartite state is 
\begin{eqnarray*}
\rho^{\prime\prime} = 
\frac{[I^{\otimes 2n} \otimes 
\Pi(\bm{f}_0,\bm{f}_1) 
\otimes I^{\otimes (4+2\theta)n}]
\rho^\prime
[I^{\otimes 2n}
 \otimes \Pi(\bm{f}_0,\bm{f}_1)
\otimes I^{\otimes (4+2\theta)n}]}
{Tr\left[
[I^{\otimes 2n} \otimes \Pi(\bm{f}_0,\bm{f}_1)
\otimes I^{\otimes (4+2\theta)n}] \rho^\prime
\right]},
\end{eqnarray*}
where
\begin{eqnarray*}
\Pi(\bm{f}_0,\bm{f}_1) &=& \sum_{\bm{g}_0,\bm{g}_1 \in \mathbf{F}_2^{\frac{n}{2}}}
\ket{\Psi_{\bm{f}_0\bm{f}_1}}\bra{\Psi_{\bm{f}_0\bm{f}_1}} \\
\ket{\Psi_{\bm{f}_0\bm{f}_1}} &=&
\frac{1}{\sqrt{2^n}} 
\sum_{\bm{l} \in \mathbf{F}_2^n}
\ket{\bm{l}} \otimes [\sigma_x^{[\bm{f}_0]}\sigma_z^{[\bm{g}_0]}
\otimes \sigma_x^{[\bm{g}_1]}\sigma_z^{[\bm{f}_1]}] \ket{\bm{l}}.
\end{eqnarray*}
Then Bob perform the recovery operation ${\cal R}_{\bm{x}\bm{z}}$, and we
have
\begin{eqnarray*}
\rho^{\prime\prime\prime} = 
[{\cal I}_n \otimes {\cal R}_{\bm{x}\bm{z}} \otimes {\cal I}_{(6+2\theta)n}]
(\rho^{\prime\prime}).
\end{eqnarray*}
We define the entanglement fidelity of the system that corresponds
to a shared key as
\begin{eqnarray}
\label{fid-conditional}
F_{\bm{x}\bm{z}|\bm{f}_0\bm{f}_1\bm{a}_T\bm{b}_T\bm{pos}\pi}
= Tr\left[
[\ket{\Phi_{\bm{x}\bm{z}}^\prime}\bra{\Phi_{\bm{x}\bm{z}}^\prime} 
\otimes I^{\otimes (6+2\theta)n}] \rho^{\prime\prime\prime}
\right].
\end{eqnarray}

To bound the mutual information between a shared key and Eve's
eavesdropping key,
we evaluate Eq.~(\ref{fid-conditional}) as follows.
Since we transmit the qubits in the $\{\ket{+},\ket{-}\}$ 
basis when $a_i=1$,
and from Eq.~(\ref{fid}), we can bound Eq.~(\ref{fid-conditional}) as 
\begin{eqnarray}
\label{fid2}
1-\mathbb{E}_{\bm{x}\bm{z}}
F_{\bm{x}\bm{z}|\bm{f}_0\bm{f}_1\bm{a}_T\bm{b}_T\bm{pos}\pi}
&\le& \sum_{(\bm{c}_{T^k},\bm{d}_{T^k}) \in {\cal E}}
{\cal P}(\bm{c}_{T^k},\bm{d}_{T^k}|\bm{f}_0,\bm{f}_1) \nonumber \\
&\le& \sum_{(\bm{c}_{T_0^k},\bm{d}_{T_1^k}) \in {\cal E}(\pi(C_1))}
{\cal P}(\bm{c}_{T_0^k},\bm{d}_{T_1^k}|\bm{f}_0,\bm{f}_1) \nonumber \\
&& + \sum_{(\bm{c}_{T_1^k},\bm{d}_{T_0^k}) \in {\cal E}(\pi(C_2^\bot))}
{\cal P}(\bm{c}_{T_1^k},\bm{d}_{T_0^k}|\bm{f}_0,\bm{f}_1),
\end{eqnarray}
where 
${\cal E}$ is the set of uncorrectable errors by the 
CSS code that is constructed by $\pi(C_1)$ and $\pi(C_2)$, and 
${\cal E}(\pi(C_1)),{\cal E}(\pi(C_2^\bot))$ are sets of 
uncorrectable errors by $\pi(C_1)$ and $\pi(C_2^\bot)$ as a part of
a CSS code respectively.
Note that $\pi(C_1)$ corrects errors caused by $\sigma_z$ and
$\pi(C_2^\bot)$ corrects errors caused by $\sigma_x$ in
qubits transmitted by the $\{\ket{+},\ket{-}\}$ basis,
and that the second inequality is due to that
the decoding error of a CSS code occurs when 
a bit flip error or a phase flip error is uncorrectable.


\subsection{Bounding the fidelity}
\label{bounding-the-fidelity}

In this section, we evaluate Eq.~(\ref{fid2}) by
taking the average of parameters, $\bm{a},\bm{b},\pi,\bm{f}_0,\bm{f}_1$.
Let fix one realization of $(\bm{c}_T,\bm{d}_T)$ and vary
$\bm{a}_T,\bm{b}_T,\bm{pos}$ at uniformly random.
Note that subsequences
$\bm{c}_{T_0^k},\bm{c}_{T_0^t},\bm{c}_{T_1^k},\bm{c}_{T_1^t}$,
$\bm{d}_{T_0^k},\bm{d}_{T_0^t},\bm{d}_{T_1^k},\bm{d}_{T_1^t}$
of $\bm{c}_T,\bm{d}_T$
vary according to $\bm{a}_T,\bm{b}_T,\bm{pos}$.
Using a lemma \cite[Lemma 5]{hamada04},
we have
\begin{eqnarray}
\Pr\left\{ |Q_{\bm{c}_{T_0^k}}(1)-Q_{\bm{c}_{T_0^t}}(1)| > \delta \ \mbox{or} \ 
	|Q_{\bm{d}_{T_1^k}}(1)-Q_{\bm{d}_{T_1^t}}(1)| > \delta \right\}
&\le& \exp \left\{ -\Theta(\delta^2 n) \right\} \nonumber \\
\Pr\left\{ |Q_{\bm{d}_{T_0^k}}(1)-Q_{\bm{d}_{T_1^t}}(1)| > \delta \ \mbox{or} \ 
	|Q_{\bm{c}_{T_1^k}}(1)-Q_{\bm{c}_{T_0^t}}(1)| > \delta \right\} && \nonumber \\
\le \exp \left\{ -\Theta(\delta^2 n) \right\}, && 
\label{prob}
\end{eqnarray}
where the base of $\exp(\cdot)$ is $2$, and
$\Theta(\delta^2 n)$ can be explicitly given as
\begin{eqnarray}
\label{theta}
\Theta(\delta^2 n) = \frac{\delta^2}{4 \ln 2}n - 2 \log (n+1) -2.
\end{eqnarray}
For each realization $\bm{c}_{T_0^t}=\bm{f}_0,\bm{d}_{T_1^t}=\bm{f}_1$,
we decide linear codes $C_1$ and 
$C_2$ that satisfy the conditions 
(\ref{con-a})--(\ref{con-c}).

First, we consider $C_1$.
Assume that $C_1$ is used over a
BSC whose crossover probability of 
first $\frac{n}{2}$ bits are $Q_{\bm{c}_{T_0^k}}(1)$
and that of latter $\frac{n}{2}$ bits are $Q_{\bm{d}_{T_1^k}}(1)$.
From the condition (\ref{con-b}), if $Q_{\bm{c}_{T_0^k}}(1) \le Q_{\bm{f}_0}(1)+\delta$
and $Q_{\bm{d}_{T_1^k}}(1) \le Q_{\bm{f}_1}(1)+\delta$,
then the decoding error probability of $C_1$ as a part of a 
CSS code is lower than or equal to $\epsilon$.
We can write the decoding error probability as
\begin{eqnarray}
\label{bsc-error}
\sum_{(P_{\zeta},P_{\eta}) \in P_{\frac{n}{2}}^2}
\epsilon_{(P_{\zeta},P_{\eta})}
Q_{BSC}(T_{(P_{\zeta},P_{\eta})}^n) =
\sum_{\bm{e} \in {\cal E}(C_1)} Q_{BSC}(\bm{e})
\le \epsilon,
\end{eqnarray}
where $Q_{BSC}(\bm{e})$ is a probability that $\bm{e}$ occurs over a 
BSC whose crossover probability of first $\frac{n}{2}$ bits are
$Q_{\bm{c}_{T_0^k}}(1)$ and that of latter $\frac{n}{2}$ bits are $Q_{\bm{d}_{T_1^k}}(1)$,
${\cal E}(C_1)$ is a set of uncorrectable errors of 
$C_1$, and
$\epsilon_{(P_{\zeta},P_{\eta})}$ is the ratio of uncorrectable
errors in $T_{(P_{\zeta},P_{\eta})}^n$, i.e.,
\begin{eqnarray*}
\epsilon_{(P_\zeta,P_\eta)} =
\frac{|T_{(P_\zeta,P_\eta)}^n \cap {\cal E}(C_1)|}{|T_{(P_\zeta,P_\eta)}^n|}.
\end{eqnarray*}
From Eq.~(\ref{bsc-error}), we have
\begin{eqnarray*}
\epsilon_{(Q_{\bm{c}_{T_0^k}},Q_{\bm{d}_{T_1^k}})}
Q_{BSC}(T_{(Q_{\bm{c}_{T_0^k}},Q_{\bm{d}_{T_1^k}})}^n) 
\le \epsilon.
\end{eqnarray*}
Using type property in  \cite[Lemma 2.6]{csiszar-korner81}, we have
\begin{eqnarray*}
Q_{BSC}(T_{(Q_{\bm{c}_{T_0^k}},Q_{\bm{d}_{T_1^k}})}^n) \ge
\frac{1}{(\frac{n}{2}+1)^2} 
2^{-n \frac{D(Q_{\bm{c}_{T_0^k}}|Q_{\bm{c}_{T_0^k}})+D(Q_{\bm{d}_{T_1^k}}|Q_{\bm{d}_{T_1^k}})}{2}}
= \frac{1}{(\frac{n}{2}+1)^2}.
\end{eqnarray*}
Thus, we have
\begin{eqnarray}
\label{uncorrectable-ratio}
\epsilon_{(Q_{\bm{c}_{T_0^k}},Q_{\bm{d}_{T_1^k}})} \le
(\frac{n}{2}+1)^2 \epsilon.
\end{eqnarray}
Consequently, if $|Q_{\bm{c}_{T_0^k}}(1)-Q_{\bm{f}_0}(1)| \le \delta$
and $|Q_{\bm{d}_{T_1^k}}(1)-Q_{\bm{f}_1}(1)| \le \delta$,
then the ratio of uncorrectable errors of $C_1$
in $T_{(Q_{\bm{c}_{T_0^k}},Q_{\bm{d}_{T_1^k}})}^n$ is less than or equal to 
$(\frac{n}{2}+1)^2 \epsilon$.

Define $J(\bm{c}_{T_0^k},\bm{d}_{T_1^k},\bm{f}_0,\bm{f}_1,C_1)$ as follows:
If $(\bm{c}_{T_0^k},\bm{d}_{T_1^k}) \in {\cal E}(C_1)$,
$J(\bm{c}_{T_0^k},\bm{d}_{T_1^k},\bm{f}_0,\bm{f}_1,C_1)=1$.
For the others, 
$J(\bm{c}_{T_0^k},\bm{d}_{T_1^k},\bm{f}_0,\bm{f}_1,C_1)=0$.
If $Q_{\bm{f}_0}(1)$ or $Q_{\bm{f}_1}(1)$ is too large and we 
abort the BB84 protocol,
then $J(\bm{c}_{T_0^k},\bm{d}_{T_1^k},\bm{f}_0,\bm{f}_1,C_1)$
is always $0$.
Note that $C_1$ is decided from $\bm{f}_0,\bm{f}_1$.
From Eq.~(\ref{prob}), $|Q_{T_0^k}(1)-Q_{\bm{f}_0}(1)| \le \delta$
and $|Q_{T_1^k}(1)-Q_{\bm{f}_1}(1)| \le \delta$ 
with high probability.
When $\bm{e},\bm{e}^\prime \in T_{(P_\zeta,P_\eta)}^n$,
there exist a permutation $\pi$ such that $\pi(\bm{e})=\bm{e}^\prime$.
Thus, if we consider the decoding error probability of $\pi(C_1)$
averaged over permutation $\pi$,
then we can consider that an error with the same type occurs with same
the probability.
We proved that if 
$|Q_{T_0^k}(1)-Q_{\bm{f}_0}(1)| \le \delta$ and
$|Q_{T_1^k}(1)-Q_{\bm{f}_1}(1)| \le \delta$,
then the ratio of uncorrectable errors of $\pi(C_1)$ in 
$T_{(Q_{\bm{c}_{T_0^k}},Q_{\bm{d}_{T_1^k}})}^n$ is less than or equal to 
$(\frac{n}{2}+1)^2 \epsilon$ in Eq.~(\ref{uncorrectable-ratio}).
Then we have 
\begin{eqnarray}
&& \mathbb{E}_{\bm{a}_T, \bm{b}_T, \bm{pos}, \pi}
J(\bm{c}_{T_0^k},\bm{d}_{T_1^k},\bm{f}_0,\bm{f}_1,\pi(C_1)) \nonumber \\
&\le& (\frac{n}{2}+1)^2 \epsilon
\Pr\left\{ |Q_{\bm{c}_{T_0^k}}(1)-Q_{\bm{c}_{T_0^t}}(1)| \le \delta \ \mbox{and} \ 
	|Q_{\bm{d}_{T_1^k}}(1)-Q_{\bm{d}_{T_1^t}}(1)| \le \delta \right\} \nonumber \\
&+& \Pr\left\{ |Q_{\bm{c}_{T_0^k}}(1)-Q_{\bm{c}_{T_0^t}}(1)| > \delta \ \mbox{or} \ 
	|Q_{\bm{d}_{T_1^k}}(1)-Q_{\bm{d}_{T_1^t}}(1)| > \delta \right\} \nonumber \\
&\le& (\frac{n}{2}+1)^2 \epsilon + \exp \left\{-\Theta(\delta^2 n)\right\}.
\label{c1bound}
\end{eqnarray}

Taking the average of Eq.~(\ref{c1bound}) over $(\bm{c}_T,\bm{d}_T)$
and exchanging the order of the averages, we have
\begin{eqnarray}
&& \mathbb{E}_{\bm{a}_T, \bm{b}_T, \bm{pos}, \pi}
\mathbb{E}_{{\cal P}(\bm{f}_0,\bm{f}_1)}
\sum_{(\bm{c}_{T_0^k},\bm{d}_{T_1^k}) \in {\cal E}(\pi(C_1))}
{\cal P}(\bm{c}_{T_0^k},\bm{d}_{T_1^k}|\bm{f}_0,\bm{f}_1) \nonumber \\
&=& \mathbb{E}_{{\cal P}(\bm{c}_T,\bm{d}_T)}
\mathbb{E}_{\bm{a}_T, \bm{b}_T, \bm{pos}, \pi}
J(\bm{c}_{T_0^k},\bm{d}_{T_1^k},\bm{f}_0,\bm{f}_1,\pi(C_1)) \nonumber \\
&\le& (\frac{n}{2}+1)^2 \epsilon + \exp \left\{-\Theta(\delta^2 n)\right\}.
\label{c1}
\end{eqnarray}
In the same way, we have 
\begin{eqnarray}
\mathbb{E}_{\bm{a}_T, \bm{b}_T, \bm{pos}, \pi}
\mathbb{E}_{{\cal P}(\bm{f}_0,\bm{f}_1)}
\sum_{(\bm{c}_{T_1^k},\bm{d}_{T_0^k}) \in {\cal E}(\pi(C_2^\bot))}
{\cal P}(\bm{c}_{T_1^k},\bm{d}_{T_0^k}|\bm{f}_0,\bm{f}_1)
\le (\frac{n}{2}+1)^2 \epsilon + \exp \left\{-\Theta(\delta^2 n)\right\}.
\label{c2}
\end{eqnarray}
From Eq.~(\ref{c1}) and (\ref{c2}),
we can rewrite Eq.~(\ref{fid2}) as
\begin{eqnarray}
1 - \mathbb{E}_{\bm{a}_T, \bm{b}_T, \bm{pos}, \pi}
\mathbb{E}_{{\cal P}(\bm{f}_0,\bm{f}_1)}
\mathbb{E}_{\bm{x}\bm{z}} 
F_{\bm{x}\bm{z}|\bm{f}_0\bm{f}_1\bm{a}_T\bm{b}_T\bm{pos}\pi}
\le 2(\frac{n}{2}+1)^2 \epsilon + 2\exp \left\{-\Theta(\delta^2 n)\right\}.
\end{eqnarray}

\subsection{Bounding the mutual information}
\label{bounding-the-mutual-information}

Using Eq.~(\ref{eve-mutual}), we can bound the mutual information as
\begin{eqnarray}
&& \rom{I}(\bm{U};\bm{E}|\bm{X},\bm{Z},\bm{A}_T,\bm{B}_T,\bm{POS},
\bm{\Pi},\bm{F}_0,\bm{F}_1) \nonumber \\
&\le& \mathbb{E}_{\bm{a}_T, \bm{b}_T, \bm{pos}, \pi}
\mathbb{E}_{{\cal P}(\bm{f}_0,\bm{f}_1)}
\rom{H}\left(
1- \mathbb{E}_{\bm{x}\bm{z}} 
F_{\bm{x}\bm{z}|\bm{f}_0\bm{f}_1\bm{a}_T\bm{b}_T\bm{pos}\pi}
\right) \nonumber \\
&& + \mathbb{E}_{\bm{a}_T, \bm{b}_T, \bm{pos}, \pi}
\mathbb{E}_{{\cal P}(\bm{f}_0,\bm{f}_1)}
(1-\mathbb{E}_{\bm{x}\bm{z}} 
F_{\bm{x}\bm{z}|\bm{f}_0\bm{f}_1\bm{a}_T\bm{b}_T\bm{pos}\pi}) 2n \nonumber \\
&\le& \rom{H}\left( 1-
\mathbb{E}_{\bm{a}_T, \bm{b}_T, \bm{pos}, \pi}
\mathbb{E}_{{\cal P}(\bm{f}_0,\bm{f}_1)}
\mathbb{E}_{\bm{x}\bm{z}} 
F_{\bm{x}\bm{z}|\bm{f}_0\bm{f}_1\bm{a}_T\bm{b}_T\bm{pos}\pi}
\right) \nonumber \\
&& + (1- \mathbb{E}_{\bm{a}_T, \bm{b}_T, \bm{pos}, \pi}
\mathbb{E}_{{\cal P}(\bm{f}_0,\bm{f}_1)}
\mathbb{E}_{\bm{x}\bm{z}} 
F_{\bm{x}\bm{z}|\bm{f}_0\bm{f}_1\bm{a}_T\bm{b}_T\bm{pos}\pi}) 2n \nonumber \\
&\le& \rom{H}\left(
2(\frac{n}{2}+1)^2 \epsilon + 2\exp \left\{-\Theta(\delta^2 n)\right\}
\right) \nonumber \\
&& + 4n(\frac{n}{2}+1)^2 \epsilon + 4n \exp\left\{-\Theta(\delta^2 n)\right\},
\end{eqnarray}
where 
$\bm{X}$, $\bm{Z}$, $\bm{A}_T$, $\bm{B}_T$,
$\bm{POS}$, $\bm{\Pi}$, $\bm{F}_0$, $\bm{F}_1$
denote the random variables of 
$\bm{x}$, $\bm{z}$, $\bm{a}_T$, $\bm{b}_T$,
$\bm{pos}$, $\pi$, $\bm{f}_0$, $\bm{f}_1$.
Using the chain rule of mutual information 
\cite[Theorem 2.5.2]{cover} and mutual 
independence of $\bm{U}$ from 
$\bm{X}$, $\bm{Z}$, $\bm{A}_T$, $\bm{B}_T$, $\bm{POS}$, $\bm{\Pi}$,
$\bm{F}_0$, $\bm{F}_1$,
we can upper bound the mutual information of the shared key
and Eve's all accessible information as
\begin{eqnarray*}
&& \rom{I}(\bm{U};\bm{E},\bm{X},\bm{A}_T,\bm{B}_T,\bm{POS},
\bm{\Pi},\bm{F}_0,\bm{F}_1) \\
&\le& \rom{I}(\bm{U};\bm{E},\bm{X},\bm{Z},\bm{A}_T,\bm{B}_T,\bm{POS},
\bm{\Pi},\bm{F}_0,\bm{F}_1) \\
&=& \rom{I}(\bm{U};\bm{E}|\bm{X},\bm{Z},\bm{A}_T,\bm{B}_T,\bm{POS},
\bm{\Pi},\bm{F}_0,\bm{F}_1) \\
&\le& \rom{H}\left(
2(\frac{n}{2}+1)^2 \epsilon + 2\exp \left\{-\Theta(\delta^2 n)\right\}
\right)
+ 4n(\frac{n}{2}+1)^2 \epsilon + 4n \exp\left\{-\Theta(\delta^2 n)\right\},
\end{eqnarray*}
where $\Theta(\delta^2 n)$ is given by Eq.~(\ref{theta}).

\section{Proof of lemma~\ref{lemma-E}}
\label{app2}

In this appendix, we prove
\begin{eqnarray*}
\min_{\scriptsize 0 \le p_0^\prime \le p_0 \atop
0 \le p_1^\prime \le p_1}
E(R,p_0^\prime,p_1^\prime)
= E(R,p_0,p_1),
\end{eqnarray*}
where we assume $p_0 < \frac{1}{2}$ and $p_1 < \frac{1}{2}$.
First, we fix $p_0^\prime$ and $p_1^\prime$ arbitrary in the 
range $0 \le p_0^\prime \le p_0$, $0 \le p_1^\prime \le p_1$,
and analyze $E(R,p_0^\prime,p_1^\prime)$ 
as a function of $R$ in Section \ref{analysis}.
Then we prove that $E(R,p_0^\prime,p_1^\prime)$ takes the
minimum at $p_0^\prime=p_0$, $p_1^\prime=p_1$ 
for arbitrary $0 \le R \le 1$ in Section \ref{minimum}.

\subsection{Analysis of $E(R,p_0^\prime,p_1^\prime)$}
\label{analysis}

Note that $p_0^\prime$ and $p_1^\prime$ are arbitrary fixed 
in the range $0 \le p_0^\prime \le p_0$,
$0 \le p_1^\prime \le p_1$ in this section.
To express $E(R,p_0^\prime,p_1^\prime)$ as a function of
$R,p_0^\prime,p_1^\prime$ explicitly,
we define
\begin{eqnarray*}
F(R,q_0,q_1) = \left\{
\begin{array}{ll}
\frac{D(q_0|p_0^\prime)+D(q_1|p_1^\prime)}{2}
+1-R-\frac{\rom{H}(q_0)+\rom{H}(q_1)}{2} & \mbox{for }
\frac{\rom{H}(q_0)+\rom{H}(q_1)}{2} < 1 - R \\
\frac{D(q_0|p_0^\prime)+D(q_1|p_1^\prime)}{2} & \mbox{for }
\frac{\rom{H}(q_0)+\rom{H}(q_1)}{2} \ge 1-R.
\end{array}
\right. 
\end{eqnarray*}
Because it is obvious that $E(R,p_0^\prime,p_1^\prime)=0$ 
for $R \ge 1-\frac{\rom{H}(p_0^\prime)+\rom{H}(p_1^\prime)}{2}$,
we assume $R < 1 - \frac{\rom{H}(p_0^\prime)+\rom{H}(p_1^\prime)}{2}$
in Section \ref{analysis}.

First, we consider the case
$\frac{\rom{H}(q_0)+\rom{H}(q_1)}{2} < 1 - R$.
If we set 
\begin{eqnarray*}
q_0^* &=& \frac{\sqrt{p_0^\prime}}{\sqrt{p_0^\prime}+\sqrt{1-p_0^\prime}}, \\
q_1^* &=& \frac{\sqrt{p_1^\prime}}{\sqrt{p_1^\prime}+\sqrt{1-p_1^\prime}},
\end{eqnarray*}
then we have 
$\left.\frac{\partial F(R,q_0,q_1^*)}{\partial q_0}\right|_{q_0=q_0^*} = 0$,
$\left.\frac{\partial F(R,q_0^*,q_1)}{\partial q_1}\right|_{q_1=q_1^*} = 0$,
and $\frac{\partial^2 F(R,q_0,q_1)}{\partial q_0^2}>0$,
$\frac{\partial^2 F(R,q_0,q_1)}{\partial q_1^2}>0$
for $0 \le q_0,q_1 \le 1$.
Thus, if $\frac{\rom{H}(q_0^*)+\rom{H}(q_1^*)}{2} < 1 - R$,
then $F(R,q_0,q_1)$ takes the minimum at $(q_0^*,q_1^*)$ and we have
\begin{eqnarray*}
E(R,p_0^\prime,p_1^\prime) =
1-R-\log\{\sqrt{p_0^\prime}+\sqrt{1-p_0^\prime}\} -
\log\{\sqrt{p_1^\prime}+\sqrt{1-p_1^\prime}\}.
\end{eqnarray*}

Next, we consider the case 
$\frac{\rom{H}(q_0^*)+\rom{H}(q_1^*)}{2} \ge 1 - R$.
$F(R,q_0,q_1)$ takes the  minimum in the range  
$\frac{\rom{H}(q_0)+\rom{H}(q_1)}{2} \ge 1 - R$.
Because we assumed  
$R < 1-\frac{\rom{H}(p_0^\prime)+\rom{H}(p_1^\prime)}{2}$,
$q_0$ and $q_1$ must be $q_0 > p_0^\prime$ or $q_1 > p_1^\prime$
in order to satisfy $\frac{\rom{H}(q_0)+\rom{H}(q_1)}{2} \ge 1 - R$.
If $q_0 < p_0^\prime$ and $q_1 > p_1^\prime$, then
$\frac{D(q_0|p_0^\prime)+D(q_1|p_1^\prime)}{2}$ can be smaller
by taking larger $q_0$ and smaller $q_1$ while 
maintaining the condition
$\frac{\rom{H}(q_0)+\rom{H}(q_1)}{2} \ge 1 - R$.
If $q_0 > p_0^\prime$ and $q_1 < p_1^\prime$, then
$\frac{D(q_0|p_0^\prime)+D(q_1|p_1^\prime)}{2}$ can be smaller
by taking larger $q_1$ and smaller $q_0$ while 
maintaining the condition
$\frac{\rom{H}(q_0)+\rom{H}(q_1)}{2} \ge 1 - R$.
Thus, $q_0$ and $q_1$ must be $q_0 \ge p_0^\prime$ and $q_1 \ge p_1^\prime$
for $\frac{D(q_0|p_0^\prime)+D(q_1|p_1^\prime)}{2}$ to be the minimum.
Thus $F(R,q_0,q_1)$ is minimum at $(q_0,q_1)$ that satisfy
$\frac{\rom{H}(q_0)+\rom{H}(q_1)}{2} = 1 - R$.
When $\frac{\rom{H}(q_0)+\rom{H}(q_1)}{2} = 1 - R$,
we can expand $\frac{D(q_0|p_0^\prime)+D(q_1|p_1^\prime)}{2}$ as
\begin{eqnarray}
\label{con-D}
&& \frac{D(q_0|p_0^\prime)+D(q_1|p_1^\prime)}{2}  \nonumber \\
&=& \frac{1}{2}\left\{
-q_0\log p_0^\prime - (1-q_0)\log (1-p_0^\prime)  \right. \nonumber \\
&& \left. - q_1\log p_1^\prime -(1-q_1)\log (1-p_1^\prime)
- \rom{H}(q_0) - \rom{H}(q_1) \right\}  \nonumber \\
&=& \frac{1}{2}\left\{
\log\frac{1}{p_0^\prime}-\log\frac{1}{(1-p_0^\prime)} \right\} q_0
+\frac{1}{2}\left\{
\log\frac{1}{p_1^\prime}-\log\frac{1}{(1-p_1^\prime)} \right\} q_1  \nonumber \\
&& + \frac{1}{2}\left\{
\log\frac{1}{(1-p_0^\prime)}+\log\frac{1}{(1-p_1^\prime)} \right\}
-(1-R).
\end{eqnarray}
Because
\begin{eqnarray*}
\frac{d \rom{H}(q_0)}{d q_0} =
\log \frac{1}{q_0} - \log \frac{1}{(1-q_0)},
\end{eqnarray*}
the tangent of the curve 
\begin{eqnarray}
\label{curve1}
\frac{\rom{H}(q_0)+\rom{H}(q_1)}{2} = 1-R
\end{eqnarray}
at $(\hat{q}_0,\hat{q}_1)$ is
\begin{eqnarray*}
\frac{1}{2}\left\{
\log\frac{1}{\hat{q}_0}-\log\frac{1}{(1-\hat{q}_0)}\right\}
(q_0-\hat{q}_0) 
+\frac{1}{2}\left\{
\log\frac{1}{\hat{q}_1}-\log\frac{1}{(1-\hat{q}_1)}\right\}
(q_1-\hat{q}_1) = 0.
\end{eqnarray*}
As is shown in Eq.~(\ref{con-D}), the curve 
\begin{eqnarray}
\label{curve2}
\frac{D(q_0|p_0^\prime)+D(q_1|p_1^\prime)}{2} = \alpha
\end{eqnarray}
is linear under the condition 
$\frac{\rom{H}(q_0)+\rom{H}(q_1)}{2} = 1 - R$,
and $\alpha$ takes the  minimum at the point where
curve~(\ref{curve1}) and line~(\ref{curve2}) touch each other
under the condition $\frac{\rom{H}(q_0)+\rom{H}(q_1)}{2} = 1 - R$.
Then, $(\hat{q}_0,\hat{q}_1)$ satisfy 
\begin{eqnarray*}
\frac{\left\{
\log\frac{1}{p_1^\prime}-\log\frac{1}{(1-p_1^\prime)} \right\}}
{\left\{
\log\frac{1}{p_0^\prime}-\log\frac{1}{(1-p_0^\prime)} \right\}}
=
\frac{\left\{
\log\frac{1}{\hat{q}_1}-\log\frac{1}{(1-\hat{q}_1)} \right\}}
{\left\{
\log\frac{1}{\hat{q}_0}-\log\frac{1}{(1-\hat{q}_0)} \right\}},
\end{eqnarray*}
and we can further rewrite
\begin{eqnarray*}
\left\{ \log\frac{1}{\hat{q}_0}-\log\frac{1}{(1-\hat{q}_0)} \right\}
&=& \beta
\left\{ \log\frac{1}{p_0^\prime}-\log\frac{1}{(1-p_0^\prime)} \right\} \nonumber \\
\left\{ \log\frac{1}{\hat{q}_1}-\log\frac{1}{(1-\hat{q}_1)} \right\}
&=& \beta
\left\{ \log\frac{1}{p_1^\prime}-\log\frac{1}{(1-p_1^\prime)} \right\}.
\end{eqnarray*}
Thus, if we set 
\begin{eqnarray*}
\hat{q}_0 &=& \frac{(p_0^\prime)^\beta}{(p_0^\prime)^\beta
+ (1-p_0^\prime)^\beta}, \\
\hat{q}_1 &=& \frac{(p_1^\prime)^\beta}{(p_1^\prime)^\beta
+ (1-p_1^\prime)^\beta}, \\
\frac{\rom{H}(\hat{q}_0)+\rom{H}(\hat{q}_1)}{2} &=& 1-R,
\end{eqnarray*}
then $F(R,q_0,q_1)$ takes the minimum at $(\hat{q}_0,\hat{q}_1)$.
Because $\frac{d\hat{q}_0}{d\beta}<0$ and 
$\frac{d\hat{q}_1}{d\beta}<0$ for $p_0^\prime<\frac{1}{2}$ and 
$p_1^\prime<\frac{1}{2}$,
$\hat{q}_0$ and $\hat{q}_1$ are decreasing functions of $\beta$.
Because we assumed 
$1-R > \frac{\rom{H}(p_0^\prime)+\rom{H}(p_1^\prime)}{2}$,
we have $\beta < 1$.
If $\beta < \frac{1}{2}$, then 
$\frac{\rom{H}(q_0^*)+\rom{H}(q_1^*)}{2}<\frac{\rom{H}(\hat{q}_0)+\rom{H}(\hat{q}_1)}{2}=1-R$
and 
$F(R,\hat{q}_0,\hat{q}_1)$ 
takes the minimum at $(q_0^*,q_1^*)$.
Thus, $\frac{1}{2} \le \beta < 1$.

Consequently, we can write $E(R,p_0^\prime,p_1^\prime)$ as
\begin{eqnarray*}
E(R,p_0^\prime,p_1^\prime)
= \left\{
\begin{array}{l}
1-R-\log\{\sqrt{p_0^\prime}+\sqrt{1-p_0^\prime}\} -
\log\{\sqrt{p_1^\prime}+\sqrt{1-p_1^\prime}\} \\
\mbox{\hspace{15mm}for } R < 1 - \frac{\rom{H}(q_0^*)+\rom{H}(q_1^*)}{2} \\
\frac{D(\hat{q_0}|p_0^\prime)+D(\hat{q_1}|p_1^\prime)}{2} \\
\mbox{\hspace{15mm}for } 1-\frac{\rom{H}(q_0^*)+\rom{H}(q_1^*)}{2} \le R 
< 1 - \frac{\rom{H}(p_0^\prime)+\rom{H}(p_1^\prime)}{2} \\
0 \\ \mbox{\hspace{15mm}for } 1 - \frac{\rom{H}(p_0^\prime)+\rom{H}(p_1^\prime)}{2} \le R.
\end{array}
\right.
\end{eqnarray*}

\subsection{The minimum of $E(R,p_0^\prime,p_1^\prime)$}
\label{minimum}

Next, we evaluate 
\begin{eqnarray*}
\min_{\scriptsize 0 \le p_0^\prime \le p_0 \atop
0 \le p_1^\prime \le p_1}
E(R,p_0^\prime,p_1^\prime)
\end{eqnarray*}
for arbitrary fixed rate $R$.
If $\frac{\rom{H}(q_0^*)+\rom{H}(q_1^*)}{2}<1-R$,
\begin{eqnarray}
\label{hani1}
E(R,p_0^\prime,p_1^\prime) = 
1-R-\log\{\sqrt{p_0^\prime}+\sqrt{1-p_0^\prime}\} -
\log\{\sqrt{p_1^\prime}+\sqrt{1-p_1^\prime}\}.
\end{eqnarray}
Eq.~(\ref{hani1}) is a decreasing function of $p_0^\prime,p_1^\prime$
for $p_0^\prime < \frac{1}{2},p_1^\prime < \frac{1}{2}$,
because if we set 
$f(p_0^\prime)=\sqrt{p_0^\prime}+\sqrt{1-p_0^\prime}$,
\begin{eqnarray*}
\frac{d f(p_0^\prime)}{d p_0^\prime} = \frac{1}{2}\left\{
\frac{1}{\sqrt{p_0^\prime}}-\frac{1}{\sqrt{1-p_0^\prime}}
\right\}
> 0
\end{eqnarray*}
for $p_0^\prime < \frac{1}{2}$ and $f(p_0^\prime)$ is
an increasing function of $p_0^\prime$.
If $\frac{\rom{H}(q_0^*)+\rom{H}(q_1^*)}{2} \ge 1-R$, then
\begin{eqnarray*}
\label{hani2}
E(R,p_0^\prime,p_1^\prime) = 
\frac{D(\hat{q_0}|p_0^\prime)+D(\hat{q_1}|p_1^\prime)}{2}.
\end{eqnarray*}
Assume $E(R,p_0^\prime,p_1^\prime)$ takes the  minimum at 
$(p_0^\prime,p_1^\prime)$ with $p_0^\prime < p_0$.
Because $p_0^\prime < \hat{q_0}$,
we define $(p_0^{\prime\prime},p_1^{\prime\prime})$
such that $p_0^\prime < p_0^{\prime\prime} < \hat{q_0}$ and
$p_1^\prime = p_1^{\prime\prime}$.
Then 
\begin{eqnarray*}
E(R,p_0^\prime,p_1^\prime) &=& 
\frac{D(\hat{q_0}|p_0^\prime)+D(\hat{q_1}|p_1^\prime)}{2} \\
&>& \frac{D(\hat{q_0}|p_0^{\prime\prime})+D(\hat{q_1}|p_1^{\prime\prime})}{2} \\
&\ge& \frac{D(\tilde{q}_0|p_0^{\prime\prime})+D(\tilde{q}_1|p_1^{\prime\prime})}{2} = E(R,p_0^{\prime\prime},p_1^{\prime\prime}),
\end{eqnarray*}
where
\begin{eqnarray*}
\tilde{q}_0 &=& \frac{(p_0^{\prime\prime})^\beta}{(p_0^{\prime\prime})^\beta
+ (1-p_0^{\prime\prime})^\beta} \\
\tilde{q}_1 &=& \frac{(p_1^{\prime\prime})^\beta}{(p_1^{\prime\prime})^\beta
+ (1-p_1^{\prime\prime})^\beta} \\
\frac{\rom{H}(\tilde{q}_0)+\rom{H}(\tilde{q}_1)}{2} &=& 1-R.
\end{eqnarray*}
Note that the first inequality is due to that
$p_0^{\prime\prime}$ is closer to $\hat{q_0}$ than 
$p_0^\prime$, and the second inequality is due to that
$(\tilde{q}_0,\tilde{q}_1)$ is the point at which
$\frac{D(q_0|p_0^{\prime\prime})+D(q_1|p_1^{\prime\prime})}{2}$ 
takes the minimum.
Thus, $E(R,p_0^\prime,p_1^\prime)$ does not take the minimum at
$(p_0^\prime,p_1^\prime)$ with $p_0^\prime < p_0$.
In a similar manner, we can show that
$E(R,p_0^\prime,p_1^\prime)$ does not take the minimum at 
$(p_0^\prime,p_1^\prime)$ with $p_1^\prime < p_1$.
Consequently, we have
\begin{eqnarray*}
\min_{\scriptsize 0 \le p_0^\prime \le p_0 \atop
0 \le p_1^\prime \le p_1}
E(R,p_0^\prime,p_1^\prime)
= E(R,p_0,p_1).
\end{eqnarray*}


\end{document}